\documentclass[manuscript,screen]{acmart}

\renewcommand\footnotetextcopyrightpermission[1]{} 
\makeatletter 
\renewcommand\@formatdoi[1]{\ignorespaces}
\makeatother

\AtBeginDocument{%
  \providecommand\BibTeX{{%
    \normalfont B\kern-0.5em{\scshape i\kern-0.25em b}\kern-0.8em\TeX}}}

\setcopyright{acmcopyright} 
\copyrightyear{2018} 
\acmYear{2018} 
\acmDOI{XXXXXXX.XXXXXXX} 

\usepackage{paralist}

\begin{document}

\title[VBL research: A Review of Video Characteristics, Tools, Technologies, and Learning Effectiveness]{A Closer Look into Recent Video-based Learning Research: A Comprehensive Review of Video Characteristics, Tools, Technologies, and Learning Effectiveness}

\author{Evelyn Navarrete}
\email{navarrete@l3s.de}
\orcid{0000-0002-5610-9908}
\affiliation{%
  \institution{L3S Research Center, Leibniz University}
  \city{Hannover}
  \country{Germany}
}

\author{Andreas Nehring}
\affiliation{%
  \institution{Institute for Science Education}
  \city{Hannover}
  \country{Germany}}
\email{nehring@idn.uni-hannover.de}
\orcid{0000-0002-8723-5552}

\author{Sascha Schanze}
\affiliation{%
  \institution{Institute of Science Education}
  \city{Hannover}
  \country{Germany}
}
\email{schanze@idn.uni-hannover.de}
\orcid{0000-0002-5570-4991}

\author{Ralph Ewerth}
\affiliation{%
 \institution{L3S Research Center, Leibniz University}
 \city{Hannover}
 \country{Germany}}
 \affiliation{%
 \institution{TIB -- Leibniz Information Centre for Science and Technology}
 \city{Hannover}
 \country{Germany}}
 \email{ralph.ewerth@tib.eu}
\orcid{0000-0003-0918-6297}

\author{Anett Hoppe}
\affiliation{%
 \institution{L3S Research Center, Leibniz University}
 \city{Hannover}
 \country{Germany}}
 \affiliation{%
 \institution{TIB -- Leibniz Information Centre for Science and Technology}
 \city{Hannover}
 \country{Germany}}
\email{anett.hoppe@tib.eu}
\orcid{0000-0002-1452-9509}

\renewcommand{\shortauthors}{Navarrete et al.}

\begin{abstract}
    People increasingly use videos on the Web as a source for learning, be it for daily tasks or in formal as well as informal educational settings.
    To support this way of learning, researchers and developers are continuously developing tools, proposing guidelines, analyzing data, and conducting experiments. 
    However, it is still not clear what characteristics a video should have to be an effective learning medium. 
    In this paper, we present a comprehensive review of 257 articles on video-based learning from computer science databases for the period from 2016 to 2021 using the PRISMA guidelines. 
    One of the aims of the review is to identify the video characteristics that have been explored by previous work.  
    Based on our analysis, we suggest a taxonomy which organizes the video
    characteristics and contextual aspects into eight categories:
    \begin{inparaenum}[(1)]
        \item audio features,
        \item visual features,
        \item textual features,
        \item instructor behavior,
        \item learners' activities,
        \item interactive features (quizzes, etc.), 
        \item production style, and
        \item instructional design.
    \end{inparaenum}
    Also, we identify four representative 
    research directions: 
    \begin{inparaenum}[(1)]
        \item proposals of tools to support video-based learning, 
        \item studies with controlled experiments, 
        \item data analysis studies, and 
        \item proposals of design guidelines for learning videos.
    \end{inparaenum}
    We find that the most explored characteristics are textual features followed by visual features, learner's activities, and 
    interactive features.
    Text of transcripts, video frames, and images (figures and illustrations) are most frequently used by tools that support learning through videos.
    The learner's activity is heavily explored through log files in data analysis studies, and 
    interactive features
    have been frequently scrutinized in controlled experiments. 
    We complement our review by contrasting research findings that investigate the impact of video characteristics on the learning effectiveness, report on tasks and technologies used to develop tools that support learning, and summarize trends in the development of design guidelines to produce learning videos.
\end{abstract}



\maketitle

\fancyfoot{} 
\thispagestyle{empty}

\section{Introduction}

Videos become increasingly a choice in the everyday life of a learner, 
facilitated not only by formal institutional platforms but also on commercial general-purpose video platforms such as YouTube.
In May 2019 it was stated that 
about 500 hours of new content was uploaded every minute~\cite{clement2019} on YouTube, where around 50\% of the views 
have been associated with learning purposes~\cite{smith2018}.
Furthermore, it was predicted that 82\% of all IP traffic in the world will be caused by video consumption by 2022~\cite{cisco2018cisco}.
Learners who turn to video sources have to rely on the technologies offered by the platform for providing suitable learning material. 
Moreover, they need to trust the expertise (if any) of the content creators and instructors 
in designing and producing effective learning videos.
However, even on dedicated platforms videos are published with no guidelines or good practices in mind~\cite{DBLP:conf/lats/GuoKR14}. 
The reason is that, although efforts to better understand video-based learning (VBL) have increased~\cite{DBLP:conf/lak/PoquetLMD18,DBLP:journals/bjet/Giannakos13, yousef2014state}, essential questions are still not answered, for example, what are the characteristics of a video\footnote{In the context of this review, we understand video characteristics as specific attributes, features, or aspects of a video that might be relevant to VBL.} to support learning in the best way, according to individual needs, the learning goals, and other contextual aspects?

Previous VBL reviews have performed descriptive analysis on the growth tendency of the field, the 
research directions, the discipline addressed by the videos, the production style, the mode of use (i.e., main or supplementary learning medium), the context or setting (i.e., formal or informal learning), video characteristics, and other dimensions~\cite{DBLP:conf/lak/PoquetLMD18, DBLP:journals/bjet/Giannakos13, yousef2014state}. 
Nevertheless, they have put special emphasis on experimental research, and did not dive into
different research directions.

In this paper, 
we present a comprehensive review of 257 articles on video-based learning from computer science databases for the period from 2016 to 2021 using the PRISMA guidelines~\cite{page_prisma_2021}.
Compared to other reviews, we have collected the largest number of articles and provide a comprehensive overview on VBL research. 
As a consequence, we can achieve a more robust analysis of the state of the art in the field.
In contrast to previous reviews that focused on experimental research, we identify and consider four main types of research articles:
\begin{inparaenum}[(1)]
    \item proposals of tools to support video-based learning, 
    \item studies with controlled experiments, 
    \item data analysis studies, and 
    \item proposals of design guidelines for learning videos.
\end{inparaenum}
We present a close investigation of
\begin{inparaenum}[(a)]
    \item video characteristics, 
    \item research approaches,
    \item tasks and technologies used to develop tools, frameworks, and pipelines that support learning,
    \item guidelines to design and create learning videos, as well as
    \item findings from controlled experiments and data analysis.
\end{inparaenum}
Furthermore, we derive a taxonomy that organizes video 
characteristics and contextual aspects into eight categories (and further sub-categories):
\begin{inparaenum}[(1)]
    \item audio features,
    \item visual features,
    \item textual features,
    \item instructor behavior,
    \item learners' activities,
    \item interactive features, 
    \item production style, and
    \item instructional design.
\end{inparaenum}
The presented results are relevant for different audiences:
\begin{inparaenum}[(a)]
    \item platform providers who wish to support their users with the best possible content,
    \item content producers who aim to develop effective learning resources,
    \item teachers who want to enhance their teaching by 
    adopting multimedia resources, and
    \item researchers who seek to update on the current state of VBL, identify potential research gaps, and bring in ideas for further research.
\end{inparaenum}

This review is an extended version of our previous work~\cite{DBLP:conf/cikm/NavarreteHE21}, where we have presented a review of the literature on video-based learning (VBL) for only two years (2020 and 2021). 
Here, we expand the reviewed period to 2016-2021 and also include the most relevant results of the articles by contrasting, analyzing and summarizing their findings. 

The rest of the paper continues with a discussion of related reviews (Section~\ref{sec:rel_wo}), the description of the methodology (Section~\ref{sec:methodology}), the analysis of the field of video-based learning, which provides an introduction to the importance of identifying the characteristics of learning videos (Section~\ref{sec:vbl}), and the results of the review (Section~\ref{sec:literature_review}). 
After that, we discuss the impact of said video characteristics on learning effectiveness (Section~\ref{sec:findings}).
Finally, we draw conclusions and point out limitations and future work (Section~\ref{sec:conclusion}).

\section{Related Work}
\label{sec:rel_wo}

\begin{table*}[ht]
\caption{Related survey articles: reference and publication year, temporal scope of the survey, number of reviewed papers, and reviewed characteristics}
\label{tab:rel_wo}
\begin{tabular}{cclll}
\toprule
\textbf{Reference} & \textbf{Review} & \textbf{N° papers} & \textbf{Reviewed dimensions} \\
 & \textbf{period} & & \\
\midrule
\cite{DBLP:journals/bjet/Giannakos13} (2013) & 2000-2012 & 166 & 1) Type of research & 4) Technology type \\ 
 & & & 2) Sample characteristics & 5) Style of use \\ 
 & & & 3) Subject area \\
\hline
\cite{yousef2014state} (2014) & 2003-2014 & 76 & 1) Learning effectiveness & 3) Design \\ 
 & & & 2) Teaching methods & 4) Reflection \\   
\hline
\cite{DBLP:conf/lak/PoquetLMD18} (2018) & 2007-2017 & 178 & 1) Video type & 3) Sample size  \\ 
& & & 2) Population & 5) Video characteristics \\  
& & & 3) Control of prior knowledge \\
\hline
\cite{sablic2020video} (2020) & 2008-2019 & 39 & Teacher perception and reflection \\
\bottomrule
\end{tabular}
\end{table*}

To the best of our knowledge, four survey articles on video-based learning have been presented previously. 
A summary with the aspects of the VBL field they have reviewed is displayed in Table~\ref{tab:rel_wo}. 
Giannakos~\cite{DBLP:journals/bjet/Giannakos13} focused increase in the number of publications on VBL, the shift of VBL research from theoretical to empirical, and the characteristics of research regarding the type, sample, discipline area, use of stable or mobile devices, and the context of the video usage (main, formal, and informal).
Yousef et al.~\cite{yousef2014state} summarized the findings of experimental research on the effectiveness of VBL, teaching methods when using videos, integration of tools with VBL content, and the use of videos to reflect on teaching methods. 
Poquet et al.~\cite{DBLP:conf/lak/PoquetLMD18} also concentrated on experimental studies. 
The authors present a descriptive analysis on the types of videos, the population to which the experiment was directed, if the experiment controlled the prior knowledge, the sample size, growth patterns in the number of publications, and the type of research (theoretical and empirical). 
In addition, the authors review the video characteristics that influence the learning effectiveness with special emphasis on the learning outcome.
The most recent work was presented by Sablic et al.~\cite{sablic2020video}, who review research on the use of videos to support teacher's self-reflection and professional development. 
There are other reviews on learning from YouTube videos 
such as the recent work by Shoufan and Mohamed~\cite{DBLP:journals/access/ShoufanM22}. 
We have not considered this study or similar studies since they did not meet our criteria of having a computer science perspective and focusing on VBL in general, and not only on a specific platform. 

All these reviews confirm the significant growth of research articles in the VBL field~\cite{DBLP:conf/lak/PoquetLMD18,DBLP:journals/bjet/Giannakos13, yousef2014state}. 
Giannakos~\cite{DBLP:journals/bjet/Giannakos13} and Poquet et al.~\cite{DBLP:conf/lak/PoquetLMD18} reported a noticeable increase of empirical and quantitative research over theoretical and qualitative research. 
Furthermore, Giannakos~\cite{DBLP:journals/bjet/Giannakos13} showed a shift in the research domain, from the social sciences to technological domains, which was also confirmed by the results of Poquet et al.~\cite{DBLP:conf/lak/PoquetLMD18} that science, technology, engineering, and mathematics (STEM) disciplines are currently a predominant research focus. 
Further important findings from Poquet et al.~\cite{DBLP:conf/lak/PoquetLMD18} suggested that the effectiveness of VBL has frequently been evaluated through two learning outcome metrics: recall (remembering information) and transfer (applying what was learned to a different scenario), which was also corroborated by Yousef et al.~\cite{yousef2014state}. 
Other popular metrics of learning effectiveness were motivation, cognitive load, mental effort, attention, and affect~\cite{DBLP:conf/lak/PoquetLMD18}.
Also, the most common sources to obtain data for analysis were clickstreams (e.g., views, likes, dislikes) and eye-tracking data. 
Regarding the video characteristics, Poquet et al.~\cite{DBLP:conf/lak/PoquetLMD18} found that the most examined characteristics were text, audio, animations, and voice over a PowerPoint presentation. 
Yousef et al.~\cite{yousef2014state} found that trending research topics were the effects of annotations and authoring tools, collaborative VBL, and video usage for reflective teacher education.

\section{Methodology}
\label{sec:methodology}

\begin{figure}[h]
\centering
\includegraphics[height=6cm]{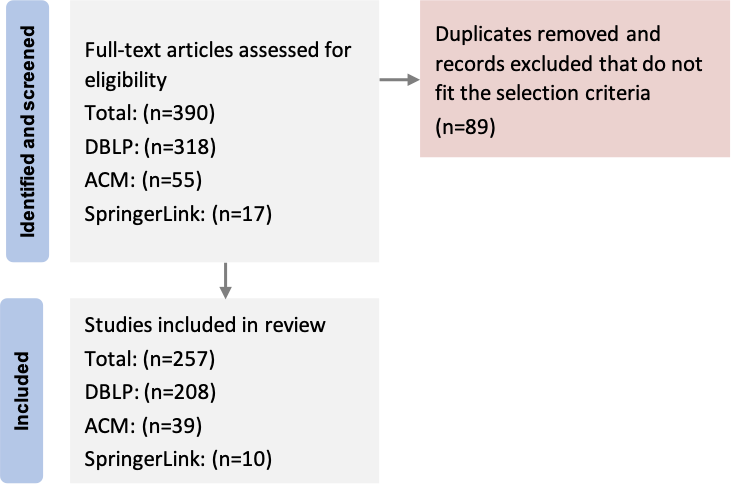}
    \caption{The diagram shows the process of finding VBL-related articles in the databases. From an initial set of 390 papers, we removed duplicates and papers that did not fit the inclusion criteria. The final set consists of 257 articles from 2016 to 2021.}
\label{prisma_framework}
\end{figure}

The methodology for our systematic review was inspired by the Preferred Reporting Items for Systematic Reviews and Meta-Analyses (PRISMA) guidelines~\cite{page_prisma_2021}. 
They provide a guide to summarize results of reviewed research articles in an appropriate manner. 

The papers covered by this review were found at SpringerLink~\cite{springer}, the Digital Library of the Association for Computing Machinery (ACM DL)~\cite{acm}, and the Digital Bibliography and Library Project (DBLP)~\cite{dblp}. 
These are focused in computer science (CS) domains, and therefore, fit well our task of reviewing the technological aspects of video-based learning.
In each of these databases, we used advanced search options as a search strategy to refine our results.
In ACM DL, we applied filters to retrieve publications that match the query terms only in the title and that fit the time frame.
In SpringerLink, we applied the same filters complemented by an additional filter to obtain articles only in CS disciplines.
In DBLP, however, we did not apply any filter since the search engine covers only CS and already retrieves articles that only match the query term in the title and sorts them all per year.

Table~\ref{tab:queries} shows the set of query combinations we used, which includes all the keywords from our previous work~\cite{DBLP:conf/cikm/NavarreteHE21} plus new terms considered for this extended review (massive open online courses \textit{MOOC} and \textit{Lecture}). We followed an iterative process where the set of query terms was updated as new terms in the field of VBL were found (please see the search order in Table~\ref{tab:queries}), and then performed a new search in the databases.

\begin{table*}[h]
\caption{This table shows the set of query terms used to find relevant articles. The base term in the first column was combined with each of the extensions in the second column (using logical AND as a connector).}
\label{tab:queries}
\begin{tabular}{lllll}
\toprule
\textbf{Base} & \textbf{Extension} & & \\
\midrule
 Video & 1.1) Teach & 1.2) Teaching & 2.1) Education & 2.2) Educational \\
 & 3.1) Instruction & 3.2) Instructional & 4.1) Learn & 4.2) Learning \\
 & 5) -based Learning & 6.1) Explanatory & 6.2) Explaining & 7) Tutorial \\ 
 & 8) Classroom & 9) Student & 10.1) Knowledge & 10.2) Knowing \\
 & 11) Skills & 12) University & 13) School & 14) Pupil  \\
 & 15) MOOC & 16) Lecture \\
\bottomrule
\end{tabular}
\end{table*}

We included a paper in the final set only when the article:
\begin{inparaenum}[(a)]
    \item considered videos in academic scenarios (this excludes video tutorials for everyday tasks such as cooking instructions),
    \item focused on videos as learning resources (this excludes the use of classroom recordings for reflective tasks such as in teacher education), and
    \item examined specific characteristics of video-based learning materials (this excludes, for instance, the use of videos to evaluate their viability in classroom).
\end{inparaenum}
Figure~\ref{prisma_framework} shows the obtained articles using the PRISMA-guided search process.
The final set of articles consists of 257 articles published in the period from 2016 to 2021.

In our data collection process, we noted the article's title, authors, year, source database, matching terms of our query set, research 
direction, research objectives, video characteristics, and, if applicable, main findings, 
discipline of the video, study sample size, video production style(s), dependent variables, and target tasks and technologies. 
All the information was assembled in a data table to facilitate efficient data access and analysis.

\section{Video-based Learning}
\label{sec:vbl}

In this section, we discuss important aspects of video-based learning, potentials, challenges, and video characteristics. This lays the groundwork and motivates our literature review in the following section.

\textbf{Potentials and Challenges:} Video-based learning is a terminology that has already been adopted in several studies (e.g.,~\cite{DBLP:journals/bjet/Giannakos13,DBLP:conf/lak/PoquetLMD18,yousef2014state, sablic2020video}), and can be considered as the task of learning using video sources, that is, acquiring knowledge and skills through video content. Videos have been implemented in classrooms as a teaching 
medium~\cite{DBLP:conf/lak/PoquetLMD18} and they form an important part of students' habits~\cite{smith2018}.
The trends show that videos might become the main learning medium in online education~\cite{hansch2015video} due to its
wide availability.
Learning platforms, such as MOOCs, can disseminate videos in simple and scalable ways~\cite{DBLP:journals/bjet/Giannakos13,DBLP:conf/lats/GuoKR14}.

The debate about the benefits of VBL continues. 
Yousef et al.~\cite{yousef2014state}, for instance, concluded from their results that learning videos are effective because they can improve the learning outcome, motivation, engagement, satisfaction, interaction, and communication between learners. 
They further argued that videos can support various learning styles and convey information otherwise hard to communicate by modalities such as plain text. 
However, the question of how to further support different learners' needs remains open, since, for example, VBL usually does not allow synchronous communication and feedback as in classic classroom settings. 

Moreover, further issues have been found in the way videos are designed and created. 
Guo et al.~\cite{DBLP:conf/lats/GuoKR14} found in their interviews to video producers 
that “video editing was not done with any specific pedagogical “design patterns” in mind” \cite[p.~45]{DBLP:conf/lats/GuoKR14} and “MOOC video producers currently base their decisions on anecdotes, folk wisdom, and best practices distilled from studies with at most dozens of subjects and hundreds of video watching sessions” \cite[p.~42]{DBLP:conf/lats/GuoKR14}. 
Additionally, methods from typical classroom scenarios have been 
transferred to create learning videos, however, it is unclear whether this can be successful~\cite{DBLP:conf/lats/GuoKR14}. 

\textbf{Examination of video characteristics:} Learning videos can be a complex learning medium since audio, visual, textual, and a large set of derived features might collide when expressing a message. 
This complexity leads us to the following research question: Which of these characteristics make a video effective for learning?
The analysis and exploration of characteristics in learning videos have been examined for different purposes, from experiments that try to understand the impact on the learning effectiveness to the construction of tools for supporting learning.
Consequently, there is a multitude of video characteristics that have been extracted, studied, and manipulated. 
These can range from low-level characteristics (e.g., text color, visual luminance, audio energy) to high-level characteristics such as models of a video's semantic content.

Video-based learning does not only involve the video and its 
instructors but also the surrounding hosting system or e-learning platform. 
Therefore, we analyzed characteristics related to the platform, for example, those related to the learner's activity, such as the usage of play and pause buttons, or number of views and likes.
Also, more complex characteristics attempt to consider the human aspects in the video (e.g., instructor gestures and speech). 
In out analysis, we focused exclusively on characteristics related to the video content and its hosting platform while excluding learner-related characteristics, as for instance, previous knowledge, age, gender, etc.

\section{Literature Review}
\label{sec:literature_review}

\begin{figure}[h]
\centering
\includegraphics[height=4cm]{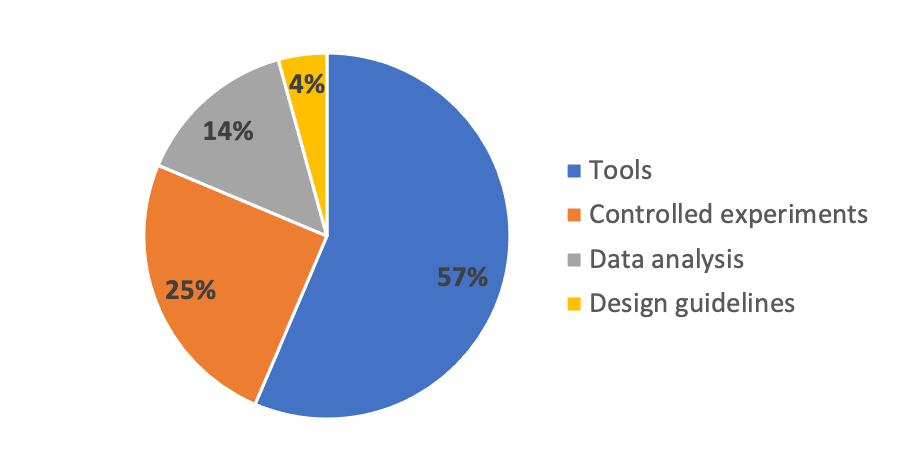}
\caption{The diagram shows the share of the different 
research directions identified in this review. 
Tools and studies that perform controlled experiments constituted the main 
research directions.}
\label{research_types}
\end{figure}

In this section, we present the results of our review and, briefly, the content of the relevant papers. 
Based on our analysis, we suggest to distinguish between four representative 
research directions:
\begin{inparaenum}[(1)]
    \item tools,
    \item controlled experiments,
    \item data analysis, and 
    \item design guidelines.
\end{inparaenum}
We found that the introduction and presentation of tools was present in more than the half of the articles, while reports about controlled experiments represented a quarter of the articles (see Figure~\ref{research_types}).

\begin{figure}[h]
\centering
\includegraphics[height=7cm]{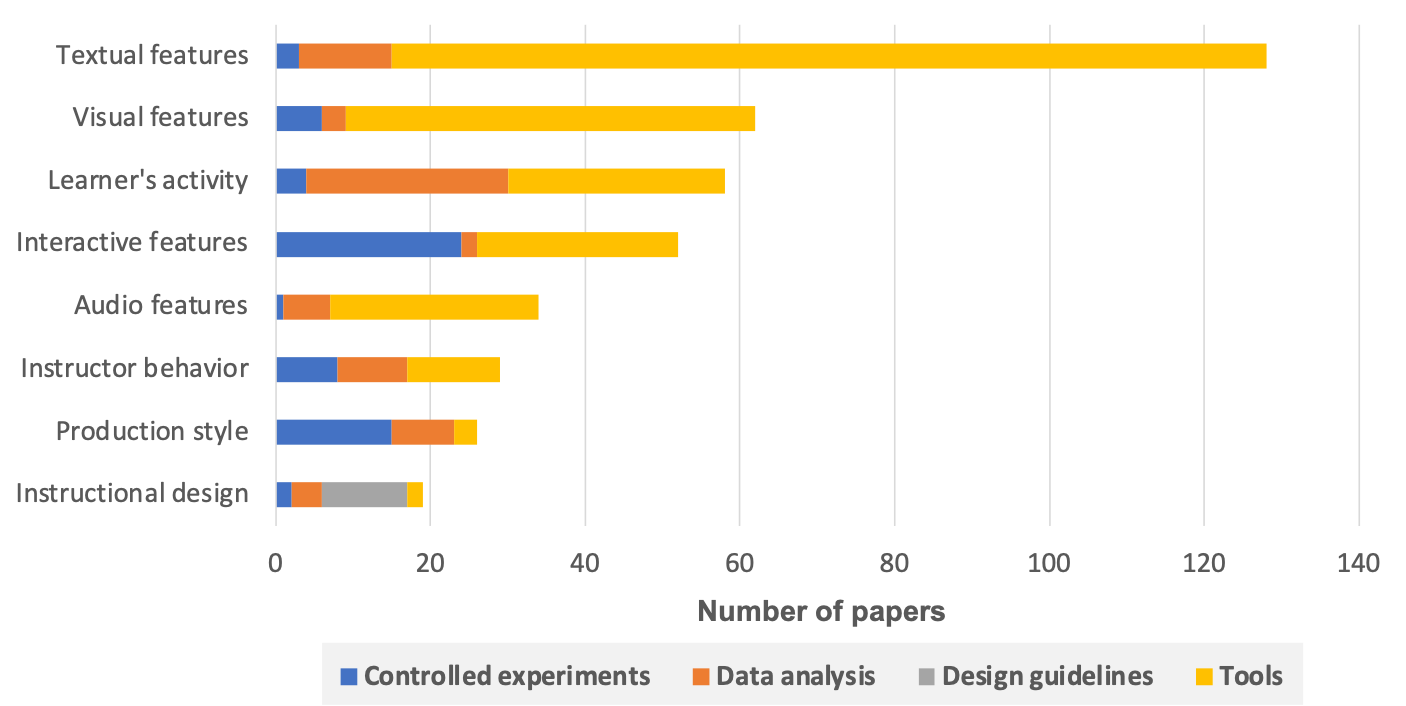}

\caption{The diagram shows the number of investigated video characteristics in the different research directions. Some video characteristics were more widely explored in certain 
research directions than in others.}
\label{features_research_type}
\end{figure}

\begin{figure}[h]
\centering
\includegraphics[height=8cm]{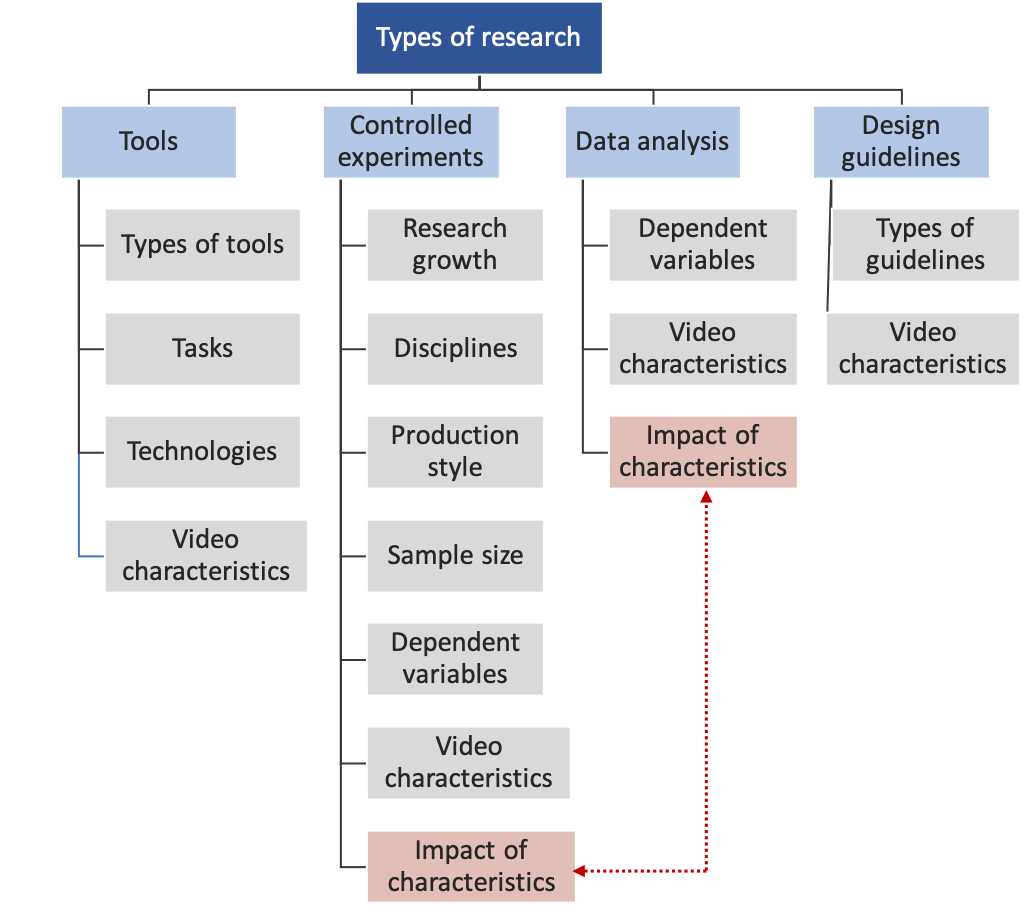}
\caption{The diagram shows the dimensions we analyzed for every type of research.}
\label{review_structure}
\end{figure}

In the following sub-sections, we report the results for each of these research directions, since this provides a natural way to analyze and contrast articles to draw conclusions.
Thus, video characteristics are discussed differently in the context of each research direction.
Table~\ref{tab:taxonomy}
summarizes the characteristics and contextual aspects identified in a single taxonomy, 
which we organized into eight main categories: \begin{inparaenum}[(1)]
    \item audio features,
    \item visual features,
    \item textual features,
    \item instructor behavior,
    \item learner's activities, 
    \item interactive features, 
    \item production style, and
    \item instructional design.
\end{inparaenum}
It is noticeable that specific characteristics fall naturally in specific categories of the taxonomy.
For example, interactive features (e.g., quizzes and annotations) have been often studied in controlled experiments but little in data analysis studies (see Figure~\ref{features_research_type}).
Figure~\ref{review_structure} depicts the structure of this section and the dimensions that were analyzed for each research type.

\begin{table*} 
\caption{Taxonomy of video characteristics}
\label{tab:taxonomy}
\begin{tabular}{p{0.082\linewidth}p{0.45\linewidth}p{0.40\linewidth}}
\toprule
\textbf{Category} & \textbf{Sub-category} & \textbf{Examples} \\
\midrule
Audio features & Audio records \cite{DBLP:journals/corr/abs-2005-13876, DBLP:conf/www/MahapatraMR18, DBLP:conf/kes/OhnishiYSNKH19, DBLP:conf/ncc/HusainM19, DBLP:conf/webmedia/SoaresB18, DBLP:journals/prl/KanadjeMAGZL16, DBLP:conf/iui/GandhiBSKLD16, DBLP:conf/iui/KumarSYD17, DBLP:conf/mm/ZhaoLLXW17, DBLP:conf/www/MahapatraMRYAR18, DBLP:conf/intellisys/BianH19, DBLP:journals/jiis/Balasubramanian16, DBLP:conf/fie/WalkYNRG20, DBLP:journals/mta/Furini18, DBLP:journals/cai/GomesDSBS19, DBLP:journals/access/LinLLCWLZL19, DBLP:conf/edm/SharmaBGPD16a, DBLP:conf/mm/ChatbriMLZKKO17, DBLP:conf/ccnc/Furini21a, DBLP:journals/csl/Sanchez-Cortina16, DBLP:conf/ism/GopalakrishnanR17}&  \\
 & Sound-related \cite{DBLP:journals/chb/Shoufan19, DBLP:journals/ijcallt/Ding18, DBLP:journals/corr/abs-2005-13876, DBLP:conf/sigite/ShoufanM17, DBLP:conf/sigir/MukherjeeTC019, DBLP:conf/webmedia/SoaresB18, DBLP:journals/prl/KanadjeMAGZL16, DBLP:conf/cikm/GhauriHE20, DBLP:journals/mta/Furini18, DBLP:conf/ictai/ImranKSK19, DBLP:conf/icrai/ChenWJ20} & Energy, entropy, spectral, jitter, shimmer, noise, Mel-frequency cepstral coefficients, perceived quality \\
 & Representation characteristics (cues) \cite{DBLP:journals/chb/WangLHS20, DBLP:journals/corr/abs-2005-13876} & Instructor-related: verbal cues \\
 
\hline

Visual features & Video frames and images 
(figures and illustrations) \cite{DBLP:journals/nca/ZhaoLQWL18, DBLP:conf/www/MahapatraMR18, DBLP:conf/ism/KokaCRSS20, DBLP:conf/ncc/HusainM19, DBLP:journals/corr/abs-1712-00575, DBLP:conf/iui/KumarSYD17, DBLP:conf/mm/ZhaoLLXW17, DBLP:conf/iui/YadavGBSSD16, DBLP:conf/www/MahapatraMRYAR18, DBLP:conf/intellisys/BianH19, DBLP:journals/nca/ZhaoLQWL18, DBLP:conf/kes/OhnishiYSNKH19, DBLP:conf/iui/YadavGBSSD16, DBLP:conf/icmcs/XuWLLZSH19, DBLP:conf/compsac/CaglieroCF19, DBLP:journals/jiis/Balasubramanian16, DBLP:conf/ism/RahmanSS20, DBLP:conf/ism/RahmanSS20, DBLP:conf/cikm/GhauriHE20, DBLP:journals/ijkesdp/KhattabiTB19, DBLP:journals/nrhm/JungSKRH19, DBLP:journals/access/DavilaXSG21, DBLP:conf/icpr/KotaSDSG20, DBLP:conf/iftc/GuanLMA19, DBLP:journals/ijdar/KotaDSSG19, DBLP:conf/icdar/XuDSG19, DBLP:conf/icfhr/DavilaZ18, DBLP:journals/mta/Furini18, DBLP:journals/mta/LeeYCC17, DBLP:conf/fie/GarberPMALS17, DBLP:conf/icig/LiuLSA17, DBLP:journals/tse/PonzanelliBMOPH19, DBLP:conf/oopsla/YadidY16, zhong2018fast, DBLP:conf/promise/AlahmadiHPHK18, DBLP:conf/lats/KhandwalaG18, DBLP:conf/icalt/CheSYM16, DBLP:journals/access/LinLLCWLZL19, DBLP:journals/nrhm/JungSKRH19, DBLP:conf/iftc/GuanLMA19, DBLP:conf/ccnc/FuriniMM18, DBLP:journals/jois/WaykarB17, DBLP:conf/softcom/CiurezMGHPJ19, DBLP:conf/icpr/XuDSG20, DBLP:conf/bigmm/GodavarthiSMR20, DBLP:conf/edm/SharmaBGPD16a, DBLP:conf/mm/WangH0YHWMW20, DBLP:conf/ccnc/Furini21a, DBLP:conf/ism/GopalakrishnanR17, DBLP:conf/icse/NongZHCZL19} & \\
 & Representation characteristics (cues) \cite{DBLP:journals/chb/WangLHS20, DBLP:journals/jche/MoonR21, DBLP:journals/ce/PiZZXYH19, DBLP:journals/jcal/WangPH19, DBLP:journals/bjet/PiHY17, DBLP:conf/icls/SharmaDGD16, DBLP:conf/www/MahapatraMR18, DBLP:conf/uist/JungSK18, DBLP:conf/icls/SharmaDGD16} & Instructor-related: pointing gestures, directed gaze. Content-related: highlighted content, arrows, moving objects \\
 & Quality-related \cite{DBLP:journals/chb/Shoufan19, DBLP:journals/ijcallt/Ding18, DBLP:conf/sigite/ShoufanM17} & video quality, resolution in pixels \\

\hline

Textual features & transcripts \cite{DBLP:journals/corr/abs-2005-13876, DBLP:conf/lats/SluisGZ16, DBLP:conf/lak/AtapattuF17, DBLP:conf/sigir/MukherjeeTC019, DBLP:journals/mta/KravvarisK19, DBLP:journals/corr/abs-2012-07589, DBLP:conf/kes/OhnishiYSNKH19, DBLP:conf/ncc/HusainM19, DBLP:conf/webmedia/SoaresB19, DBLP:conf/mmm/GalanopoulosM19, DBLP:conf/webmedia/SoaresB18, DBLP:conf/iui/GandhiBSKLD16, DBLP:conf/iui/KumarSYD17, DBLP:conf/mm/ZhaoLLXW17, DBLP:conf/iui/YadavGBSSD16, DBLP:conf/www/MahapatraMR18, DBLP:conf/intellisys/BianH19, ma2019automatic, DBLP:conf/iui/YooJLJ21, DBLP:conf/t4e/VimalakshaVK19, DBLP:journals/jiis/Balasubramanian16, DBLP:journals/jasis/KimK16, DBLP:conf/cikm/GhauriHE20, DBLP:journals/corr/abs-1807-03179, DBLP:conf/lats/JoYK19, DBLP:journals/nrhm/JungSKRH19, DBLP:conf/fie/WalkYNRG20, DBLP:journals/mta/KravvarisK19, DBLP:journals/mta/Furini18, DBLP:journals/kais/AlbahrCA21, DBLP:journals/cai/GomesDSBS19, DBLP:journals/tse/PonzanelliBMOPH19, DBLP:conf/icse/Escobar-AvilaPH17, DBLP:conf/icalt/NangiKRB19, DBLP:conf/icalt/CheSYM16, DBLP:conf/icalt/TavakoliHEK20, DBLP:conf/ideal/BleoancaHPJM20, DBLP:conf/ectel/SchultenMLH20, DBLP:conf/paams/JordanVTB20, DBLP:conf/chi/ZhaoBCS18, DBLP:conf/edm/AytekinRS20, DBLP:conf/uist/JungSK18, DBLP:conf/www/TangLWSL21, DBLP:conf/fie/WalkYNRG20, DBLP:conf/icalt/BorgesR19, DBLP:journals/snam/KravvarisK17, DBLP:conf/mmm/BasuYSZ16, DBLP:conf/mir/CooperZBS18, DBLP:conf/dms/CoccoliV16, DBLP:conf/icuimc/JiKJH18, DBLP:conf/chi/FraserNDK19, DBLP:conf/hais/StoicaHPJM19, DBLP:conf/cbmi/BasuYZ16, DBLP:conf/esws/DessiFMR17, DBLP:conf/mm/ChatbriMLZKKO17, DBLP:conf/mm/WangH0YHWMW20, DBLP:conf/tale/HorovitzO20, DBLP:conf/specom/HernandezY21} &  \\
 & Metadata \cite{DBLP:conf/helmeto/EradzeDFP20, DBLP:journals/ce/SaurabhG19, DBLP:journals/chb/Shoufan19, DBLP:journals/oir/LeeOGK17, silva2017instructional, DBLP:journals/i-jep/HildebrandA18, DBLP:conf/ectel/SjodenDM18, DBLP:journals/tele/Meseguer-Martinez19, DBLP:conf/icse/GalsterMG18, DBLP:journals/corr/abs-2012-07589, DBLP:journals/corr/abs-1807-03179, DBLP:journals/mta/KravvarisK19, DBLP:journals/cai/GomesDSBS19, DBLP:journals/tse/PonzanelliBMOPH19, DBLP:conf/icse/Escobar-AvilaPH17, DBLP:conf/icalt/CheSYM16, DBLP:conf/icalt/TavakoliHEK20, DBLP:conf/paams/JordanVTB20, DBLP:conf/chi/ZhaoBCS18, DBLP:conf/www/TangLWSL21, DBLP:conf/mmm/BasuYSZ16, DBLP:conf/mir/CooperZBS18, DBLP:journals/access/BorgesS19, DBLP:conf/chi/FraserNDK19, DBLP:conf/aied/StepanekD17, DBLP:conf/cist/OthmanAJ16, DBLP:conf/chi/SungHSCLW16, DBLP:conf/hais/StoicaHPJM19, DBLP:journals/eait/MubarakCA21, DBLP:conf/edm/BulathwelaPLYS20, DBLP:conf/cbmi/BasuYZ16, DBLP:conf/iwpc/PocheJWSVM17, DBLP:conf/bdca/OthmanAJ17, DBLP:journals/ijmlo/LeiYLTYL19} & Titles, keywords, tags, video length, comments, course syllabus, video url \\
 & On-screen text \cite{DBLP:journals/zmp/ZeeAPSG17, DBLP:journals/ets/OzdemirIS16, DBLP:conf/sigcse/WhitneyD19, DBLP:journals/nca/ZhaoLQWL18, DBLP:conf/www/MahapatraMR18, DBLP:conf/kes/OhnishiYSNKH19, DBLP:conf/ncc/HusainM19, DBLP:journals/corr/abs-2012-07589, DBLP:conf/iui/GandhiBSKLD16, DBLP:conf/iui/KumarSYD17, DBLP:conf/mm/ZhaoLLXW17, DBLP:conf/www/MahapatraMR18, DBLP:conf/ism/KokaCRSS20, DBLP:conf/intellisys/BianH19, DBLP:conf/icmcs/XuWLLZSH19, DBLP:conf/compsac/CaglieroCF19, DBLP:journals/jiis/Balasubramanian16, DBLP:journals/access/DavilaXSG21, DBLP:conf/icpr/KotaSDSG20, DBLP:conf/icpr/KotaSDSG20, DBLP:conf/icdar/XuDSG19, DBLP:conf/icfhr/DavilaZ18, DBLP:journals/mta/LeeYCC17, DBLP:conf/fie/GarberPMALS17, DBLP:journals/tse/PonzanelliBMOPH19, DBLP:conf/oopsla/YadidY16, zhong2018fast, DBLP:conf/promise/AlahmadiHPHK18, DBLP:journals/corr/abs-1807-03179, DBLP:conf/lats/KhandwalaG18, DBLP:conf/icalt/CheSYM16, DBLP:conf/ccnc/FuriniMM18, DBLP:journals/jois/WaykarB17, DBLP:conf/softcom/CiurezMGHPJ19, DBLP:conf/esws/DessiFMR17, DBLP:conf/icse/NongZHCZL19} & \\
 & Representation characteristics \cite{DBLP:conf/www/MahapatraMR18}  &  Font size, boldeness \\

\hline

Instructor behavior & Body-related \cite{DBLP:journals/ce/BeegeNSNSWMR20, DBLP:journals/chb/BeegeNSR19, DBLP:journals/bjet/WangLCWS19, DBLP:journals/chb/HorovitzM21, DBLP:journals/chb/LiKBJ16, DBLP:journals/ijcallt/Ding18, DBLP:conf/ecce/TianB16, DBLP:conf/icrai/ChenWJ20, DBLP:conf/icpr/XuDSG20, DBLP:conf/bigmm/GodavarthiSMR20, DBLP:journals/chb/HoogerheideWNG18, DBLP:conf/sigite/ShoufanM17} & Gestures: hand strokes and waving. Facial emotions, body language, pose, clothes, gender appearance\\
 & Speech-related \cite{DBLP:journals/chb/HorovitzM21, DBLP:journals/chb/Shoufan19, DBLP:journals/ijcallt/Ding18, DBLP:conf/lats/SluisGZ16, DBLP:conf/lak/AtapattuF17, DBLP:conf/sigite/ShoufanM17, DBLP:conf/cchi/PatelGZ18, DBLP:conf/www/MahapatraMR18, DBLP:conf/webmedia/SoaresB19, DBLP:journals/jiis/Balasubramanian16, DBLP:conf/lats/JoYK19, DBLP:journals/nrhm/JungSKRH19, DBLP:conf/icalt/CheSYM16, DBLP:journals/nrhm/JungSKRH19, DBLP:conf/icuimc/JiKJH18, DBLP:conf/edm/BulathwelaPLYS20} & Tone, speech rate, intonation, lexical diversity, syllable count, prosodic characteristics, readability, silence rate \\
 & Attitudes \cite{DBLP:journals/chb/Shoufan19} &  Enthusiasm \\
 
\hline

Learner's activity & 
 \cite{DBLP:conf/helmeto/EradzeDFP20, DBLP:journals/bjet/StohrSMNM19, DBLP:conf/amia/WynnWLSBWC19, DBLP:journals/jche/CostleyFLB21, DBLP:conf/sigcse/AngraveZHM20, DBLP:journals/ijhci/Li19, DBLP:conf/educon/RodriguezNFPM18, DBLP:journals/itse/CostleyL17, DBLP:journals/bjet/GiannakosJK16, DBLP:journals/nca/HuZGW20, DBLP:journals/aiedu/UchidiunoKHYO18, DBLP:conf/lats/Kovacs16, DBLP:conf/lats/SluisGZ16, DBLP:journals/ce/Shoufan19, DBLP:conf/lak/LeiGHOQKYL17, DBLP:journals/elearn/YanB18, DBLP:conf/chiir/DodsonRFYHF18, DBLP:conf/lak/AtapattuF17, DBLP:conf/cchi/PatelGZ18, DBLP:conf/ieeevast/HeZD18, DBLP:conf/fie/WalkYNRG20, DBLP:conf/icetc/ChoiHHLHPS19, DBLP:conf/fie/LongTS18, DBLP:journals/access/ZhangYCL18, DBLP:conf/lwmoocs/BotheM20a, DBLP:conf/lak/WachtlerKTE16, DBLP:conf/edm/MbouzaoDS19, DBLP:journals/jcp/MinCX19, DBLP:conf/icalt/HeZDY18, DBLP:conf/www/Zhang17b, DBLP:journals/tsp/BrintonBCP16, DBLP:journals/eait/MubarakCA21, DBLP:journals/corr/abs-2002-01955, DBLP:conf/chi/LeeM20, DBLP:conf/wecwis/AouifiEHMD21, DBLP:conf/goodit/FuriniGM21, DBLP:conf/aied/MbouzaoDS20, DBLP:journals/eait/LemayD20, DBLP:conf/cits/KorosiEFT18, DBLP:conf/bdca/OthmanAJ17, DBLP:conf/aied/LalleC20,
 DBLP:journals/ce/SaurabhG19, DBLP:journals/chb/Shoufan19, DBLP:journals/ijmlo/LeiYLTYL19,  DBLP:journals/tele/Meseguer-Martinez19, DBLP:journals/oir/LeeOGK17, DBLP:conf/sigite/ShoufanM17, DBLP:conf/icalt/TavakoliHEK20, DBLP:journals/snam/KravvarisK17, DBLP:conf/icccsec/ZhangZWL18, DBLP:conf/www/TangLWSL21,
 DBLP:journals/chb/MerktBFS18, DBLP:journals/tkl/SchroederCC20, DBLP:journals/bjet/Garrett21, DBLP:conf/lak/LangCMP20, DBLP:conf/edm/BulathwelaPLYS20} & 
 Explicit: play, pause, speed rate, seek forward and backward. Implicit: viewing percentage, average of videos watched per week.
Popularity-related: number of views, likes, dislikes, user ratings.
Pace-related: automatic pauses, pauses controlled by the learner, type of pauses, fast pace. \\

\hline

Interactive features & \cite{DBLP:conf/educon/KleftodimosE18, DBLP:journals/ce/AltenPJK20, DBLP:conf/hci/LeisnerZRC20, DBLP:conf/hicss/KellerLFL19, DBLP:conf/lak/ZeeDSGGSPA18, DBLP:journals/ijcses/SozeriK21, DBLP:conf/ectel/TaskinHHDM19, DBLP:conf/cscwd/LuLW21, DBLP:journals/ijicte/KuhailA20, chen2021exploring, DBLP:conf/ecis/WinklerHFSSL20, DBLP:journals/sle/WachtlerHZE16, DBLP:journals/chb/CojeanJ17, DBLP:conf/iiaiaai/KuYC19, DBLP:journals/eait/PalaigeorgiouP19, DBLP:journals/tkl/MoosB16, shelton2016exploring, DBLP:conf/sac/PimentelYMZ19, DBLP:journals/bjet/Chen20, DBLP:conf/ilrn/Munoz-CarpioCB20, DBLP:journals/ce/Araiza-AlbaKMSS21, DBLP:journals/access/DaherS21, DBLP:journals/bjet/HuangHC20,DBLP:conf/hvei/GuervosRPMDG19, DBLP:journals/ijcallt/Ding18, DBLP:journals/caee/GranjoR18, DBLP:conf/jcdl/FongDZRF18, DBLP:conf/chi/LiuYWW19, DBLP:journals/sle/HeraultLMFE18, DBLP:conf/fie/WalkYNRG20, DBLP:journals/ile/LinC19, DBLP:conf/educon/MalchowBM18, DBLP:conf/uist/JungSK18, DBLP:conf/ihci/DebPB17, DBLP:conf/csee/JoL17, DBLP:conf/chi/NguyenL16, DBLP:conf/lak/SharmaAJD16, DBLP:journals/sle/Kohen-VacsMRJ16, DBLP:conf/lak/KleftodimosE16, DBLP:journals/tse/BaoXXL19, DBLP:conf/hci/ToTT16, DBLP:conf/icbl/ZhuPS17, llanda2019video, DBLP:conf/chi/LiuKW18, lai2018study, DBLP:conf/sigcse/SinghAML16, DBLP:conf/lats/KhandwalaG18, DBLP:conf/ht/RobalZLH18, DBLP:conf/lak/WachtlerKTE16, DBLP:conf/vrst/KoumaditisC19, DBLP:conf/teem/BernsMRD18, DBLP:journals/ijai/WachtlerE19} & Quizzes embedded in the video, virtual reality and 360-degree characteristics, in-video annotations, feedback to the learner, games, hyperlinks, interface elements for navigation \\ 

\hline

Production style & \cite{DBLP:conf/icetc/CaoNW18, ohashi2019comparison, DBLP:journals/ile/HewL20, DBLP:journals/jcal/PiHY17, DBLP:journals/chb/StullFM18, DBLP:conf/chi/SrivastavaVLEB19, DBLP:journals/access/DavilaXSG21, DBLP:conf/lak/SrivastavaNLVE020, DBLP:conf/chi/LeeM20, DBLP:conf/collabtech/NugrahaWZHI20, DBLP:journals/mta/Perez-NavarroGC21, DBLP:journals/ijim/Al-KhateebA20, DBLP:journals/ijicte/CooksonKH20, DBLP:conf/icls/DingASBC18, DBLP:conf/aied/StrancM19, DBLP:journals/ijcallt/Ding18, DBLP:journals/itse/CostleyL17, DBLP:journals/i-jep/HildebrandA18, rahim2019video, DBLP:conf/lats/ThorntonRW17, DBLP:journals/chb/Shoufan19, DBLP:journals/ijmlo/LeiYLTYL19, DBLP:conf/sigite/ShoufanM17, DBLP:conf/edm/SharmaBGPD16a, DBLP:conf/iva/HartholtRFM20, DBLP:conf/gcce/HayashiBN20a} & Khan-style, with a talking head, dialogue-style or monologue-style, slide-based, with animations, paper-based explanations, classroom-style videos,  \\
   
\hline

Instructional design and principles & \cite{DBLP:conf/hci/Zhang21, DBLP:journals/bjet/Meij19, DBLP:conf/helmeto/EradzeDFP20, DBLP:journals/chb/Shoufan19, DBLP:conf/lats/OuGJH16, DBLP:journals/ijmlo/LeiYLTYL19, DBLP:journals/or/Acuna-SotoLP20, DBLP:conf/icccsec/ZhangZWL18, DBLP:conf/sigcse/MinnesAGF19, DBLP:journals/bjet/WijnkerBGD19, DBLP:conf/icis/WeinertGBB20, DBLP:conf/iceit/GeL20, DBLP:conf/tale/LeiYKLA16, DBLP:conf/icteri/Voronkin19, lange2020improving, DBLP:conf/sigcse/Stephenson19, DBLP:conf/lats/DodsonRFYHF18, hodges2018ensuring, DBLP:journals/ce/HungKC18} & Multimedia learning principles, guidelines for effective engaging, pedagogical characteristics of didactical mathematical knowledge, according to film theory, interaction theory, limited capacity theory   \\
 
\hline

Others & \cite{DBLP:conf/sigcse/SchreiberD17, DBLP:conf/icchp/KikusawaOKR16, DBLP:journals/jcal/MeijM16, DBLP:journals/zmp/ZeeAPSG17, DBLP:journals/corr/abs-2005-13876} & Music, sign language, inclusion of a summary, content design, visual complexity, content design: text layout, image position, slide size, slide design, text ration, image ration \\
\bottomrule
\end{tabular}
\end{table*}

\subsection{Tools, Pipelines, Frameworks, and Methods}
\label{subsec: tools}

\begin{figure}[h]
\centering
\includegraphics[height=4cm]{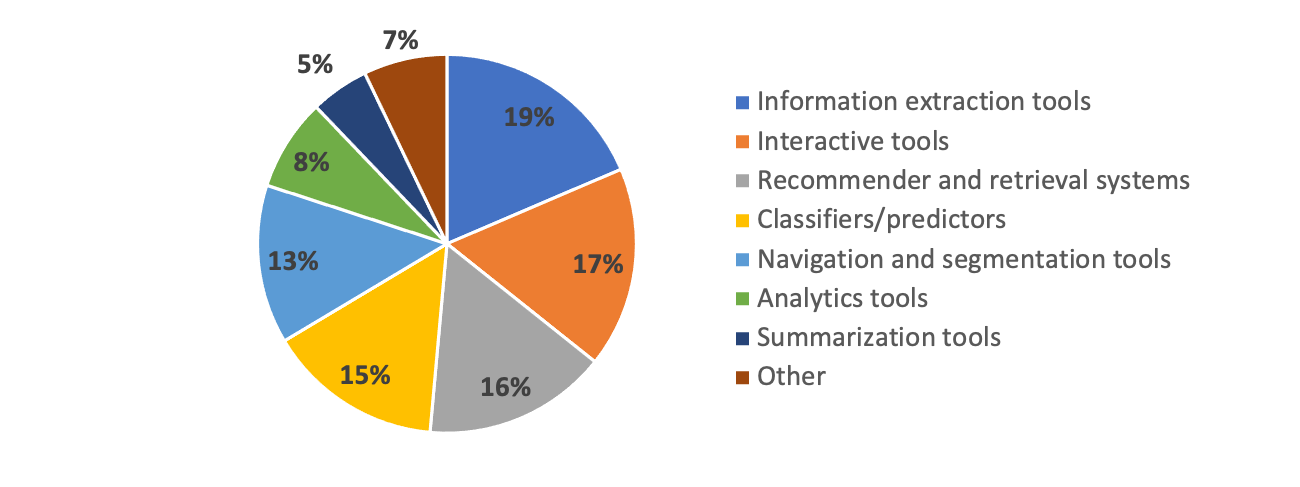}
\caption{types of tools that support video-based learning identified in this review}
\label{tool_types}
\end{figure}

In this section, we give an overview of the tools that the research community developed to assist and enhance video-based learning. 
We also report the automatic tasks carried out by the tools and the technologies that supported it. 

We identified and classified the articles according to the following types of tools:
\begin{inparaenum}[(a)]
    \item information extraction methods, 
    \item interactive tools, 
    \item recommendation systems, 
    \item classifiers, 
    \item video navigation and segmentation tools, 
    \item video analysis tools and methods, 
    \item video summarization, and
    \item others, 
\end{inparaenum}
where the first five types represented about 80\% of all developed tools (see Figure~\ref{tool_types}).
Also, please note that some articles proposed systems consisting of two or more tools~\cite{DBLP:conf/uist/JungSK18, DBLP:conf/fie/WalkYNRG20,  DBLP:conf/lak/WachtlerKTE16, DBLP:conf/lats/KhandwalaG18, DBLP:conf/www/Zhang17b}.

\begin{figure}[h]
\centering
\includegraphics[height=8.5cm]{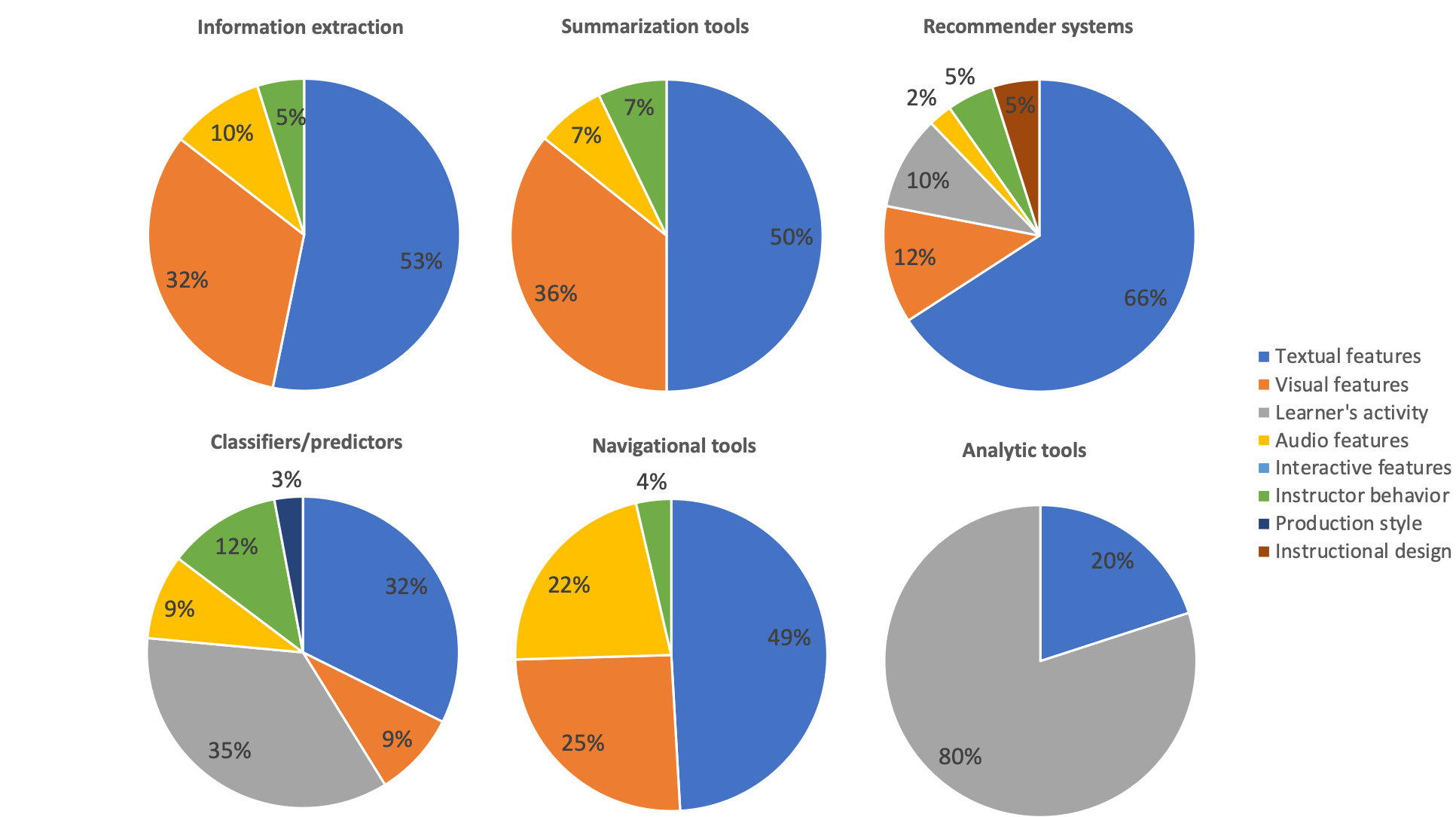}
\caption{The diagram displays the shares for targeted video characteristics used to develop tools for video-based learning. 
Textual and visual features were the mostly used characteristics, mainly by recommendation systems, information extraction, summarization, and navigational tools.}
\label{features_tools}
\end{figure}

Figure~\ref{features_tools} shows the video characteristics that have been used in articles to develop VBL tools. 
It can be seen that text and visual features were the most used characteristics, mainly by recommender systems, information extraction and summarization methods, as well as navigational tools.  
Regarding tasks and technologies, we identified some 
low-level tasks performed as basic steps, of which, many rely on deep learning approaches.
For example, for textual video elements, several studies employed optical character recognition~\cite{DBLP:conf/icmcs/XuWLLZSH19, DBLP:conf/www/MahapatraMR18, DBLP:journals/corr/abs-1807-03179, DBLP:journals/access/DavilaXSG21, DBLP:conf/intellisys/BianH19, DBLP:conf/compsac/CaglieroCF19, DBLP:journals/ijdar/KotaDSSG19, DBLP:conf/ncc/HusainM19, DBLP:conf/icdar/XuDSG19, DBLP:conf/icfhr/DavilaZ18, DBLP:conf/ccnc/FuriniMM18, DBLP:journals/mta/Furini18, DBLP:journals/mta/LeeYCC17, DBLP:journals/jois/WaykarB17, DBLP:conf/icalt/CheSYM16, DBLP:journals/jiis/Balasubramanian16, DBLP:conf/esws/DessiFMR17, DBLP:conf/mm/ZhaoLLXW17, DBLP:journals/cai/GomesDSBS19, DBLP:journals/tse/PonzanelliBMOPH19, DBLP:conf/oopsla/YadidY16, zhong2018fast, DBLP:conf/www/MahapatraMRYAR18, DBLP:conf/icse/NongZHCZL19, DBLP:conf/lats/KhandwalaG18}, keyword extraction~\cite{DBLP:conf/specom/HernandezY21, DBLP:conf/ism/KokaCRSS20, DBLP:journals/mta/KravvarisK19, DBLP:conf/csee/JoL17, DBLP:journals/jois/WaykarB17, DBLP:journals/prl/KanadjeMAGZL16, DBLP:journals/jiis/Balasubramanian16, DBLP:conf/hci/Zhang21, DBLP:conf/mir/CooperZBS18, DBLP:conf/dms/CoccoliV16, DBLP:conf/iui/KumarSYD17, DBLP:journals/cai/GomesDSBS19, DBLP:conf/icse/Escobar-AvilaPH17, DBLP:conf/icuimc/JiKJH18, DBLP:conf/iui/YadavGBSSD16, DBLP:conf/www/MahapatraMRYAR18, DBLP:conf/ectel/SchultenMLH20}, generic natural language processing methods (e.g.,~\cite{DBLP:conf/compsac/CaglieroCF19, DBLP:journals/mta/KravvarisK19, DBLP:journals/snam/KravvarisK17,DBLP:conf/lats/SluisGZ16, DBLP:conf/iwpc/PocheJWSVM17}), or utilized word embeddings (e.g.,~\cite{DBLP:conf/cikm/GhauriHE20, DBLP:conf/specom/HernandezY21, DBLP:conf/icalt/NangiKRB19, DBLP:conf/mmm/GalanopoulosM19, DBLP:conf/esws/DessiFMR17, DBLP:conf/hci/ToTT16}). 
For image-related data, many studies suggested approaches for slide transition detection and video segmentation~\cite{DBLP:conf/www/MahapatraMR18, DBLP:journals/ijkesdp/KhattabiTB19, DBLP:journals/tkl/SchroederCC20, DBLP:conf/uist/JungSK18, DBLP:conf/iftc/GuanLMA19, DBLP:conf/compsac/CaglieroCF19, DBLP:conf/ccnc/FuriniMM18, DBLP:journals/mta/Furini18, DBLP:conf/fie/GarberPMALS17, DBLP:conf/icig/LiuLSA17, DBLP:conf/icalt/CheSYM16, DBLP:conf/chi/NguyenL16, DBLP:conf/mm/ZhaoLLXW17, DBLP:conf/mmm/GalanopoulosM19, DBLP:conf/webmedia/SoaresB18, DBLP:conf/cbmi/BasuYZ16, DBLP:conf/iui/GandhiBSKLD16, DBLP:conf/oopsla/YadidY16, ghosh2021augmenting, DBLP:conf/icdar/XuDSG19}.
For audio tracks, various studies carried out speech recognition to generate transcripts~\cite{DBLP:conf/hci/LeisnerZRC20, DBLP:conf/fie/WalkYNRG20, DBLP:conf/iui/YooJLJ21, DBLP:conf/intellisys/BianH19, DBLP:conf/kes/OhnishiYSNKH19, DBLP:conf/compsac/CaglieroCF19, DBLP:conf/icalt/BorgesR19, DBLP:conf/ncc/HusainM19, DBLP:conf/mmm/GalanopoulosM19, DBLP:journals/mta/Furini18, DBLP:conf/csee/JoL17, DBLP:journals/csl/Sanchez-Cortina16, DBLP:journals/jiis/Balasubramanian16, DBLP:conf/mmm/BasuYSZ16, DBLP:conf/esws/DessiFMR17, DBLP:conf/dms/CoccoliV16, DBLP:conf/iui/GandhiBSKLD16, DBLP:conf/mm/ZhaoLLXW17, DBLP:journals/cai/GomesDSBS19, DBLP:conf/www/MahapatraMRYAR18, DBLP:conf/mm/ChatbriMLZKKO17}.
In the next subsections, we discuss the tasks and technologies in detail according to each type of tool. 
Within each subsection, we have grouped the tool proposals according to their purpose and the video characteristics that they have employed.

\subsubsection{Information Extraction Tools}
\label{subsec: info extraction}

The reviewed articles in this section focused mainly on the extraction of text from videos (handwritten or machine printed) and metadata, as well as 
the extraction of key information such as the extraction of key video segments.

\textbf{Extraction of text:}
We found seven proposals that extracted handwritten text~\cite{DBLP:journals/access/DavilaXSG21, DBLP:conf/icpr/KotaSDSG20, DBLP:journals/ijdar/KotaDSSG19, DBLP:conf/icdar/XuDSG19, DBLP:journals/mta/LeeYCC17, DBLP:conf/fie/GarberPMALS17} and two papers on extracting machine-printed text (e.g., from slide-based videos)~\cite{zhong2018fast, DBLP:journals/ijkesdp/KhattabiTB19}. 
Davila et al.~\cite{DBLP:journals/access/DavilaXSG21} retrieved handwriting using a convolutional neural network (CNN). 
Kota et al.~\cite{DBLP:journals/ijkesdp/KhattabiTB19} used multiscale histogram of gradients for feature representation and deep neural networks to extract handwriting. 
In another article, Kota et al.~\cite{DBLP:journals/ijdar/KotaDSSG19} detected bounding boxes with handwriting and used 
clustering algorithms to group them, which could identify also content that was occluded by the instructor when she was standing in front of the whiteboard. 
The handwriting was then extracted by enclosing it in bounding boxes and binarizing it. 
Lee et al.~\cite{DBLP:journals/mta/LeeYCC17} detected changes 
in the the handwriting pixels and extracted them using an adaptive threshold and time-series denoising techniques.
Davila and Zanibbi~\cite{DBLP:conf/icfhr/DavilaZ18} extracted handwritten formulas using a random forest classifier trained on handwritten symbols.
Xu et al.~\cite{DBLP:conf/icdar/XuDSG19} identified the main instructor's actions such as writing and erasing as clues to identify the video content. 
The actions were detected using pose estimation technologies. 
In the same way, Garber et al.~\cite{DBLP:conf/fie/GarberPMALS17} identified handwriting through instructor's actions of writing and erasing. 
The instructor could be removed by masking the regions where there is a change among video frames.
Zhong and Ma~\cite{zhong2018fast} extracted machine printed text using the maximally stable extremal regions method to detect candidate regions and clustering techniques were used to ensemble lines of text. 
Finally, Khattabi et al.~\cite{DBLP:journals/ijkesdp/KhattabiTB19} detected parts of the video that contained text using an AdaBoost meta-classifier.

\textbf{Extraction of key information:} Ghauri et al.~\cite{DBLP:conf/cikm/GhauriHE20} classified key segments in videos using different fusion schemes and bi-long-short-term memory network modules. 
Che et al.~\cite{DBLP:conf/icalt/CheSYM16} detected the most important video segments using the slides structure and speech characteristics. 
Kravvaris and Kermanidis~\cite{DBLP:journals/mta/KravvarisK19} identified important topics in the transcripts using Latent Dirichlet Allocation. 
They contrasted them with relevant terms from the learners' comments which were found using the Maximum Likelihood algorithm. 
Lin et al.~\cite{DBLP:journals/access/LinLLCWLZL19} classified knowledge from non-knowledge by using a CNN that processes image and sound simultaneously. 
The classifier is trained on public datasets containing audio and visual human conversations, cooking events, and also collected video lectures; however, it was not clarified what the criteria were to classify something as "knowledge". 
Similarly, Liu et al.~\cite{DBLP:journals/corr/abs-1807-03179} extracted medical information according to their level of knowledge (low or high) by building a Logistic Regression classifier using the text from transcripts.
Finally, for computer programming video tutorials, Ponzanelli et al.~\cite{DBLP:journals/tse/PonzanelliBMOPH19} classified segments according to the kind of content, for example, theoretical concepts, software source code, etc. 
The classifier employs temporal, structural, semantic, and code-related characteristics.

\textbf{Extraction of metadata:} 
Jung et al.~\cite{DBLP:journals/nrhm/JungSKRH19} extracted keywords from transcripts using the TextRank algorithm. Tags were extracted using singular value decomposition techniques and other relevant information was extracted using entity recognition techniques.
Escobar-Avila et al.~\cite{DBLP:conf/icse/Escobar-AvilaPH17} compared three algorithms (term frequency–inverse document frequency, Latent Dirichlet Allocation using genetic algorithms, and the BM25F ranking function) to find the best technique to extract tags from video titles, description, and transcripts.
Gomes et al.~\cite{DBLP:journals/cai/GomesDSBS19} extracted related topics with the help of a knowledge base, where a knowledge graph was built measuring the semantic similarity between concepts. 
The similarity was based on the number of resources (Uniform Resource Identifiers of instances in the knowledge base) the videos share and the categories that the resources share, and was calculated with the Sorensen-Dice coefficient.
Furini~\cite{DBLP:journals/mta/Furini18} extracted tags by measuring the most frequent words.
Ghosh et al.~\cite{ghosh2021augmenting} detected 
concepts that are not familiar or do not belong to the domain of the video content 
but could be relevant to better understand the video.
The concepts were linked to Wikipedia articles, which were used to build a concept similarity network. 
Unfamiliar topics were extracted using community detection techniques.
Similarly, Nangi et al.~\cite{DBLP:conf/icalt/NangiKRB19} extracted unfamiliar topics, which were considered as cross-discipline concepts, also using community analysis in a concept graph. 

\textbf{Other:} Part of the studies presented tools to extract key-phrases~\cite{DBLP:conf/oopsla/YadidY16, DBLP:journals/kais/AlbahrCA21}, slides~\cite{DBLP:conf/iftc/GuanLMA19, DBLP:conf/icig/LiuLSA17}, and 
software source code (from video tutorials to learn computer programming)~\cite{DBLP:conf/lats/KhandwalaG18, DBLP:conf/oopsla/YadidY16, DBLP:conf/promise/AlahmadiHPHK18}.

\subsubsection{Interactive Tools}
\label{subsec: interactive tools}

The development of interactive tools (tools that implement features allowing for communication and interaction between the video and the learner) mainly targeted in-video annotations, quizzes embedded in the video, and, interestingly, feedback to the learner for preventing mind wandering.

\textbf{In-video Annotations:}
These tools offered a feature that allowed the learner to take notes while watching the video, and also were accompanied with other supporting characteristics.
Fong et al.~\cite{DBLP:conf/jcdl/FongDZRF18} suggested a tool in which the learner can interact not only through annotations but also interface elements for navigation.
Liu et al.~\cite{DBLP:conf/chi/LiuYWW19} allowed the learner to create annotations and highlight parts of the transcript. 
Malchow et al.~\cite{DBLP:conf/educon/MalchowBM18} let the learner create annotations for each slide presented in the video. 
Lai et al.~\cite{lai2018study} did not only offer text-based annotations but also sketch-based annotations where the learner could draw 
into the video frame. 
Similarly, Singh et al.~\cite{DBLP:conf/sigcse/SinghAML16} created a tool for text-based and sketch-based annotations, and supported the student in comment creation, color-coding, and 
interface elements for navigation.
Nguyen and Liu~\cite{DBLP:conf/chi/NguyenL16} automated the note taking process by tracking eye gaze.
The tracking system automatically extracted relevant parts as candidate notes. 
Deb et al.~\cite{DBLP:conf/ihci/DebPB17} allowed the viewer to create annotations but did not introduce another supporting feature. 

\textbf{Quizzes:} Herault et al.~\cite{DBLP:journals/sle/HeraultLMFE18} proposed a 360-degree interactive environment with quizzes embedded in the video and 
feedback to support the learner. 
Kohen-Vacs et al.~\cite{DBLP:journals/sle/Kohen-VacsMRJ16} introduced a tool to embed quizzes and announcements along the playback into the video.
Walk et al.~\cite{DBLP:conf/fie/WalkYNRG20} included quizzes in the video with the  aim of enhancing active learning.
Wachtler et al.~\cite{DBLP:conf/lak/WachtlerKTE16} presented an application that can include several types of questions such as single- and multiple-choice questions, as well as questions to the instructor. 
Ma and Ma~\cite{ma2019automatic} developed an automatic question generation method using the video transcript and an external 
knowledge graph.
Wachtler and Ebner~\cite{DBLP:journals/ijai/WachtlerE19} created an application to schedule quizzes and other interactive features along the playing video.

\textbf{
Feedback to learner for preventing mind wandering:} Lin et al.~\cite{DBLP:journals/ile/LinC19} introduced feedback in learning videos to alert the viewers in mind-wandering states. 
The tool kept track of brainwave signals to detect loss of focus and recommended to review again parts of the video.
Jo and Lim~\cite{DBLP:conf/csee/JoL17} instead used a game that identified high-frequency words. 
To detect mind wandering, the game consisted in presenting a word to the learner and ask if that word was present previously in the video. 
Robal et al.~\cite{DBLP:conf/ht/RobalZLH18} made use of web page, mouse, and face tracking technologies to determine whether the viewer was paying attention. 
If the user was inattentive, the system rang a bell, displayed a red frame around the video, and paused it.
Sharma et al.~\cite{DBLP:conf/lak/SharmaAJD16} employed eye gaze tracking to identify mind wandering and displayed red marks when distraction was detected. 
In a similar way, Llanda~\cite{DBLP:conf/iciit/Llanda19} offered a tool that provided feedback to the learner, although, not with regard to mind wandering but to the sentiment state. 
The tool tracked the viewers' emotions by analyzing face patterns and detecting negative sentiment to consequently suggest re-watching the video.

\textbf{Other:} Some studies suggested tools to create videos that integrate 360-degree characteristics~\cite{DBLP:journals/sle/HeraultLMFE18, DBLP:conf/vrst/KoumaditisC19, DBLP:conf/teem/BernsMRD18}. 
Other papers introduced clickable objects~\cite{DBLP:conf/uist/JungSK18, DBLP:conf/hci/ToTT16}, interactive concept maps~\cite{DBLP:conf/chi/LiuKW18}, feedback for the learners adapted to their previously identified 
proficiency level~\cite{DBLP:conf/icbl/ZhuPS17}, and tools to interact directly with the video content for practicing, for example, image editing~\cite{DBLP:conf/lak/KleftodimosE16} and computer programming~\cite{DBLP:journals/tse/BaoXXL19, DBLP:conf/lats/KhandwalaG18}.

\subsubsection{Recommender and Retrieval Systems}
\label{subsec:recommender systems}

Recommender systems have mainly considered the similarity between transcripts and metadata to suggest new videos.
There were, however, other interesting characteristics that have been exploited, for instance, video popularity (e.g., likes and views), learning style, teaching style, etc. 

\textbf{Based on the similarity of transcripts and metadata:} Zhao et al.~\cite{DBLP:conf/chi/ZhaoBCS18} suggested recommendations based on topic similarity, whereby the video is represented by means of its latent topics using the Latent Dirichlet Allocation algorithm.
Waykar and Bharathi~\cite{DBLP:journals/jois/WaykarB17} retrieved videos based on keywords extracted from on-screen text and also the texture of keyframes using a texture histogram. 
Then, the videos were arranged with the help of Probability Extended Neighbor Classification. 
Bleoanca et al.~\cite{DBLP:conf/ideal/BleoancaHPJM20} retrieved the top-N videos based on topic similarity using Latent Semantic Indexing to extract them, clustering techniques to group videos by discipline, and Wikipedia articles to augment the query context. 
Basu et al.~\cite{DBLP:conf/mmm/BasuYSZ16} recommended videos when the learner was navigating Wikipedia pages. 
They used the Latent Dirichlet Allocation algorithm to extract relevant topics from Wikipedia content and video transcripts.
Jung et al.~\cite{DBLP:journals/nrhm/JungSKRH19} recommended similar videos based on keywords, tags, and the readability of the video.
In a similar manner, Ji et al.~\cite{DBLP:conf/icuimc/JiKJH18} generated recommendations using metrics of linguistic difficulty and similarity of keywords identified by the TextRank algorithm. 
Jordán et al.~\cite{DBLP:conf/paams/JordanVTB20} built a system that considers not only the characteristics of the current video but also previous videos watched by the learner.
Walk et al.~\cite{DBLP:conf/fie/WalkYNRG20} made use of word embeddings constructed from the video transcript and compared them against the user search query to retrieve the most similar video segment. 
Tavakoli et al.~\cite{DBLP:conf/icalt/TavakoliHEK20} built a system to recommend videos for learning skills required in the labor market. 
The authors use the skills that can be learned from the video and the skills demanded by the market together with video popularity features (e.g., likes and views) and length.

\textbf{Based on ontologies and external knowledge bases:} Schulten et al.~\cite{DBLP:conf/ectel/SchultenMLH20} determined related concepts through external ontologies based on Wikipedia articles and extracted keywords using the term frequency–inverse document frequency algorithm to discover new learning resources 
through a web search API.
In the same way, Borges and Reis~\cite{DBLP:conf/icalt/BorgesR19} extracted concepts and their relationships from an existing knowledge base to recommend videos. 
Coccoli and Vercelli~\cite{DBLP:conf/dms/CoccoliV16} extracted concepts and also related concepts by linking the content to Wikipedia articles. 
The learners could receive recommendations of videos and also Web pages.
Borges and Silveira~\cite{DBLP:journals/access/BorgesS19} 
retrieved videos by analyzing the subject-object-predicate construction of the user query and searching keywords in a semantic graphic database. 

\textbf{Playlists:} Aytekin et al.~\cite{DBLP:conf/edm/AytekinRS20} assembled a set of videos for recommendations based on prerequisite relationship detection techniques, that is, identifying when a concept refers to another concept. 
Likewise, Tang et al.~\cite{DBLP:conf/www/TangLWSL21} recommended a playlist by building a concept map that showed the most relevant concepts in videos and how they are linked to each other. 
The authors also considered the sentiment analyzed from likes, dislikes, and comments.
Furini et al.~\cite{DBLP:conf/ccnc/FuriniMM18}  recommended a playlist composed of video segments that covered the keywords given by the learner. 
For segmenting the video, silence periods and also visual transitions were detected using histogram-based techniques. 
Techniques for optical character recognition were employed to understand the video content and match it to the user query. 
Cooper et al.~\cite{DBLP:conf/mir/CooperZBS18} recommended videos based on the main topics, which were found by using the Latent Dirichlet Allocation algorithm and inter-topic relationships from the course syllabus. 
For creating the playlist, the authors used sequential pattern mining techniques to discover relationships.

\textbf{Other:} Further studies presented recommender systems based on video popularity~\cite{DBLP:journals/snam/KravvarisK17, DBLP:conf/icccsec/ZhangZWL18}, videos delivered by the same instructor~\cite{DBLP:conf/iftc/GuanLMA19, DBLP:conf/ictai/ImranKSK19}, viewer's learning style~\cite{DBLP:conf/softcom/CiurezMGHPJ19}, the current software application that the learner is using~\cite{DBLP:conf/chi/FraserNDK19}, and multiple pedagogical criteria~\cite{DBLP:journals/or/Acuna-SotoLP20}.

\subsubsection{Classifiers and Predictors}
\label{subsec: classifiers}

Classifiers and predictors
were mainly used to estimate learning outcome, while some other noticeable tasks were discipline classification, instructor behavior prediction, and dropout prediction.

\textbf{Learning outcome prediction:} Lemay and Doleck~\cite{DBLP:journals/eait/LemayD20} used a repeated incremental pruning algorithm to predict the learners' grades based on the frequency of video viewing.
Mbouzao et al.~\cite{DBLP:conf/aied/MbouzaoDS20} predicted whether a learner will pass or fail a MOOC course using attendance rate (number of videos the student played divided by the number of videos mandatory to watch), 
utilization rate (percentage of video play time activity divided by the sum of video lengths of all videos), and watching index (attendance rate multiplied by utilization rate) metrics in a straightforward rule that compared a single student against the average.
Mubarak et al.~\cite{DBLP:journals/eait/MubarakCA21} trained a long-short-term memory network to predict the learning outcome as pass or fail based on learner's activity data (e.g., play, pause, speed, etc.). 
Furukawa et al.~\cite{DBLP:conf/iiaiaai/FurukawaIY20} used the number of views, interval of video views (e.g., 0 to 19 days), and the first access to videos in a multiple regression predictor to evaluate whether the learner will pass or fail a MOOC course. 
Similarly, Korosi et al.~\cite{DBLP:conf/cits/KorosiEFT18} predicted whether a learner will pass or fail a MOOC course using random forests and bagging based on characteristics related to platform usage (e.g., scrolling actions) and not representing the video itself. 
In a different way to determine learning success, Lallé and Conati~\cite{DBLP:conf/aied/LalleC20} clustered learners into low and high achievers using
characteristics such as the average proportion of videos watched per week.
Aouifi et al.~\cite{DBLP:conf/wecwis/AouifiEHMD21}, instead, predicted learning outcome in terms of passing or failing the next quiz. 
The authors take as characteristics the sequence of previous quiz results to feed them into a recurrent neural network. 
Brinton et al. ~\cite{DBLP:journals/tsp/BrintonBCP16} also predicted whether the learner will pass a quiz successfully but using
characteristics such as play and pause, revisiting, etc.

\textbf{Discipline classification:} Stoica et al.~\cite{DBLP:conf/hais/StoicaHPJM19} built a support vector machine trained on a Wikipedia dataset to classify video transcripts and keywords into the discipline to which they belong. 
Dessi et al.~\cite{DBLP:conf/esws/DessiFMR17} introduced a model based on variants of support vector machine algorithms to classify videos into their corresponding disciplines using transcripts, on-screen text features, and a relevance score for each feature. 
Basu et al.~\cite{DBLP:conf/cbmi/BasuYZ16} clustered videos according to their disciplines by analyzing the topic similarity between the transcript and Wikipedia articles.
Othman et al.~\cite{DBLP:conf/bdca/OthmanAJ17} classified videos into categories such as Education and Science and Technology but also to rather different types including How-to and Style.
The authors used decision trees and a Naive Bayes classifier for classification based on a various video metadata features.
In a different approach, Chatbri et al.~\cite{DBLP:conf/mm/ChatbriMLZKKO17} built a CNN fed with image-like representations that characterize the transcript to then classify the video into its discipline. 
The images are generated using statistical co-occurrence transformation techniques, which arranges the transcript into an array of characters and uses ASCII codes of the characters to create gray scale images.

\textbf{Instructor behavior prediction:} Chen et al.~\cite{DBLP:conf/icrai/ChenWJ20} built a neural network to predict teachers' enthusiasm using facial expressions, voice, volume, and pose characteristics. 
Godavarthi~\cite{DBLP:conf/bigmm/GodavarthiSMR20} also used facial landmarks characteristics but to predict instructor emotions using a CNN.
Sharma et al.~\cite{DBLP:conf/edm/SharmaBGPD16a} predicted video liveliness (energetic or active) using two different long-short-term memory networks fed with audio features, visual features from the instructor actions, and production setup (e.g., instructor sitting or standing, slide-based video or whiteboard-based video).
Xu et al.~\cite{DBLP:conf/icpr/XuDSG20} predicted the instructor's actions (write, pick eraser, erase, etc.) based on pose characteristics using a random forest and adaptive CNN classifiers.

\textbf{Dropout prediction:} Jeon and Park~\cite{DBLP:journals/corr/abs-2002-01955} used
learner's activity data to predict weekly course dropout in a MOOC platform by embedding click data sequences and feeding them into a gate recurrent unit. 
Furini et al.~\cite{DBLP:conf/goodit/FuriniGM21} investigated dropout prediction in an e-learning platform relying on characteristics related to video access, and using random forest and K-nearest neighbor classification. 

\textbf{Other:} Further articles proposed models to predict how engaging a video was~\cite{DBLP:conf/edm/BulathwelaPLYS20}, when a learner will jump back~\cite{DBLP:conf/www/Zhang17b}, and to classify learners' comments~\cite{DBLP:conf/iwpc/PocheJWSVM17}.

\subsubsection{Navigation and Segmentation Tools}
\label{subsec: navigation systems}

Many articles proposed navigation in the style of a table of contents, but also navigation using keywords and visual elements has been presented. 
The suggested tools commonly use segmentation techniques to match the extracted information with the corresponding video segment to allow user navigation.

\textbf{Table of contents navigation:} Such tools consist of a table of contents which the user can click on and jump to the corresponding part in the video. 
These approaches usually allow users to find segments where a certain sub-topic or keyword is presented.
The segments to which the sub-topic or keyword belong are identified and indexed to allow the learner to click on them and to jump there directly~\cite{DBLP:journals/corr/abs-2012-07589, DBLP:journals/mta/KravvarisK19, DBLP:conf/www/MahapatraMR18, DBLP:conf/sigir/MukherjeeTC019, DBLP:conf/ncc/HusainM19, DBLP:conf/iui/GandhiBSKLD16, DBLP:conf/iui/KumarSYD17, DBLP:conf/mm/ZhaoLLXW17, DBLP:conf/iui/YadavGBSSD16, DBLP:conf/intellisys/BianH19}.
Sub-topics and keywords have been extracted with technologies such as the Latent Dirichlet Allocation algorithm~\cite{DBLP:conf/sigir/MukherjeeTC019, DBLP:journals/mta/KravvarisK19, DBLP:conf/ncc/HusainM19, DBLP:conf/intellisys/BianH19}, TextRank algorithm~\cite{DBLP:conf/iui/YadavGBSSD16, DBLP:conf/www/MahapatraMRYAR18}, Markov chain Monte Carlo method~\cite{DBLP:conf/intellisys/BianH19}, and concept extraction algorithms~\cite{DBLP:conf/www/MahapatraMRYAR18}.

\textbf{Keyword navigation:} Another frequent kind of proposal for navigation extracted keywords and indexed them according to specific segments of the video~\cite{DBLP:conf/kes/OhnishiYSNKH19, DBLP:conf/ism/KokaCRSS20, DBLP:journals/mta/KravvarisK19, DBLP:journals/prl/KanadjeMAGZL16, DBLP:conf/iui/KumarSYD17, DBLP:conf/mm/ZhaoLLXW17}. 
 
\textbf{Navigation using visual elements:} This kind of approach enabled exploration by presenting visual parts of the video, for example, main slides, figures, tables, charts, etc., which the viewer clicked and then jumped to the corresponding segment in the video~\cite{DBLP:journals/nca/ZhaoLQWL18, DBLP:journals/corr/abs-1712-00575, DBLP:conf/iui/KumarSYD17, DBLP:conf/mm/ZhaoLLXW17, DBLP:conf/iui/YadavGBSSD16}.

\textbf{Video segmentation:} One of the natural ways of splitting a video is to detect slide changes, 
which can be used to extract key video frames. 
For video segmentation, TextTiling~\cite{DBLP:conf/sigir/MukherjeeTC019, DBLP:conf/mm/ZhaoLLXW17} and dynamic programming~\cite{DBLP:conf/iui/GandhiBSKLD16} were suggested.
For the detection of slide changes, edge-based differences between video frames~\cite{DBLP:conf/mm/ZhaoLLXW17} and overlapping patches~\cite{DBLP:conf/www/MahapatraMR18} were expoited. 
Also, we found studies dedicated to exclusively find better segmentation methods for learning videos (e.g.\cite{DBLP:conf/webmedia/SoaresB19, DBLP:conf/mmm/GalanopoulosM19, DBLP:conf/webmedia/SoaresB18}). 
These approaches used natural language processing for phrase extraction, named entity recognition, and part-of-speech-tagging~\cite{DBLP:conf/mmm/GalanopoulosM19} to segment the video.
Other studies used audio features to segment according to sound characteristics (e.g., power spectral density, fundamental frequencies, low level acoustic characteristics)~\cite{DBLP:conf/webmedia/SoaresB18, DBLP:conf/webmedia/SoaresB19}.

\subsubsection{Analytic Tools}
\label{subsec: analytic tools}

Analytic tools relied heavily in 
the learner's activity data to understand video usage, and in some cases, contrasted this usage with the learners' performance.
Another small group of analytic tools used characteristics such as metadata and eye gaze.

\textbf{Based on learner's activity:} 
General analytic tools exploited log data of learner activity from e-learning platforms to show how videos are used.
Walk et al.~\cite{DBLP:conf/fie/WalkYNRG20} developed a tool that shows the audience retention for specific video segments, the number of times the segment was re-watched, and how much time learners spent on it.
Wachtler et al.~\cite{DBLP:conf/lak/WachtlerKTE16} presented a tool that provided an analysis about the video (e.g., number of views) and also about the student (e.g., extent to which the video was watched). 
Long et al.~\cite{DBLP:conf/fie/LongTS18} proposed a framework to analyze patterns regarding video usage, for instance, usage of a single video, specific video segments, all videos in a course, etc.

Other tools used log data to understand the learners' behavior in relation to their academic performance. 
Min et al.~\cite{DBLP:journals/jcp/MinCX19} suggested an analytic system that contrasted the log data and the student performance on the quizzes with the watching behavior.
He et al.~\cite{DBLP:conf/ieeevast/HeZD18} analyzed the viewing behavior using two metrics, where the first one records if the learner watched the video and the second one records the amount of video viewing. 
Later, He at al.~\cite{DBLP:conf/icalt/HeZDY18} showed how these metrics can be adopted to understand the results in academic performance.

In addition, 
learner activity data has been used in other interesting analytic tools for diverse purposes. 
Mbouzao et al.~\cite{DBLP:conf/edm/MbouzaoDS19} presented a method to embed the viewer's sequence of activities into a vector and found if they characterize a specific video, this is, if a certain video elicits particular behavior. 
Bothe and Meinel~\cite{DBLP:conf/lwmoocs/BotheM20a} developed a metric to track viewers' visits to previous MOOC videos, considering that learners do not always strictly follow the path suggested by a platform.  
Choi et al.~\cite{DBLP:conf/icetc/ChoiHHLHPS19} included smartwatch data in their system log data to improve the measurement of interest and difficulty of a video.

\textbf{Based on video metadata:} Othman et al.~\cite{DBLP:conf/cist/OthmanAJ16}  
mined MOOC videos by exploiting their metadata (e.g., video length, comments, title, country).
Stepanek and Dorn~\cite{DBLP:conf/aied/StepanekD17} presented a framework to analyze viewers' comments to identify and predict the engagement generated by a video.
Sung et al.~\cite{DBLP:conf/chi/SungHSCLW16} analyzed viewers' comments considering the sentiment, its relevance, its type, and its topic for displaying this information in the video timeline.

\textbf{Other:} Zhang et al.~\cite{DBLP:journals/access/ZhangYCL18}  
analyzed learning videos using data from eye tracking systems.

\subsubsection{Summarization Tools}
\label{subsec: summarization tools}

Summarization  
approaches aim at generating a concise and shorter 
representation of the entire video. 
Text summarization was the most applied approach, however, also summarization of visual elements was suggested.

\textbf{Text-based summarization:} This kind of summarization was achieved by sentence segmentation, named entity recognition,  
extraction of key phrase extract, subtitles, and other metadata using transcripts as a main source~\cite{DBLP:conf/t4e/VimalakshaVK19, DBLP:journals/jiis/Balasubramanian16, DBLP:journals/jasis/KimK16} and on-screen text in a few cases~\cite{DBLP:conf/icmcs/XuWLLZSH19, DBLP:conf/compsac/CaglieroCF19}. 
Methods that supported this task were term frequency–inverse document frequency~\cite{DBLP:conf/t4e/VimalakshaVK19}, Naive Bayes~\cite{DBLP:journals/jiis/Balasubramanian16}, and Latent Semantic Analysis~\cite{DBLP:conf/iui/YooJLJ21} for identifying the relevant information.

\textbf{Image-based summarization:} Rahman et al.~\cite{DBLP:conf/ism/RahmanSS20} proposed a visual summary by extracting a group of 
relevant images (figures or illustrations) from slide-based videos. 
The image was included or not in the summary according to its importance, which was computed based on its size, 
unique aspects in the image, and the duration of its occurrence in the video.

\subsubsection{Other}

Further interesting articles proposed methods to create videos automatically
\cite{DBLP:conf/iva/HartholtRFM20, DBLP:conf/gcce/HayashiBN20a, DBLP:conf/ccnc/Furini21a}, alleviate the split attention problem~\cite{DBLP:conf/icls/SharmaDGD16, DBLP:conf/ism/GopalakrishnanR17}, enhance speech recognition~\cite{DBLP:conf/specom/HernandezY21, DBLP:journals/csl/Sanchez-Cortina16}, and find content that is similar not in other videos but in other media such as in textbooks~\cite{DBLP:conf/mm/WangH0YHWMW20} and books~\cite{DBLP:conf/tale/HorovitzO20}.

\subsection{Controlled Experiments}

In this section, we report general findings regarding the evolution of experimental research, main disciplines, production style, and sample sizes.  We also present the video characteristics that have been tested in controlled experiments 
and the dependent variables used to measure the impact of those characteristics (e.g., on engagement, learning outcome, etc.).

\subsubsection{
Trends, Disciplines, Production Style, and Sample Size}

\textbf{Trends:} There is an increase of articles on controlled experiment for the last three years in comparison to the previous three years. 
More than 60\% of articles in this category have been published in the last three years~\cite{DBLP:journals/ce/BeegeNSNSWMR20, DBLP:journals/ce/AltenPJK20, DBLP:conf/hci/LeisnerZRC20, DBLP:journals/bjet/Chen20, DBLP:conf/hicss/KellerLFL19, DBLP:conf/aied/StrancM19, DBLP:journals/chb/WangLHS20, DBLP:journals/tkl/SchroederCC20, DBLP:journals/jche/MoonR21, DBLP:journals/ijcses/SozeriK21, DBLP:conf/ilrn/Munoz-CarpioCB20, DBLP:journals/ijicte/CooksonKH20, DBLP:journals/ijim/Al-KhateebA20, DBLP:journals/ce/Araiza-AlbaKMSS21, DBLP:conf/ectel/TaskinHHDM19, DBLP:journals/access/DaherS21, DBLP:journals/mta/Perez-NavarroGC21, DBLP:conf/cscwd/LuLW21, DBLP:conf/collabtech/NugrahaWZHI20, DBLP:journals/bjet/WangLCWS19, DBLP:journals/ijicte/KuhailA20, DBLP:journals/bjet/HuangHC20, DBLP:journals/bjet/Garrett21, DBLP:conf/chi/LeeM20, DBLP:conf/lak/LangCMP20, chen2021exploring, DBLP:conf/lak/SrivastavaNLVE020, DBLP:journals/chb/HorovitzM21, DBLP:journals/access/DavilaXSG21, DBLP:conf/ecis/WinklerHFSSL20, DBLP:journals/ce/PiZZXYH19, DBLP:conf/chi/SrivastavaVLEB19, DBLP:conf/hci/Zhang21,  DBLP:journals/chb/HoogerheideWNG18, DBLP:conf/iiaiaai/KuYC19, DBLP:conf/hvei/GuervosRPMDG19, DBLP:conf/sac/PimentelYMZ19, DBLP:conf/sigcse/WhitneyD19, ohashi2019comparison, DBLP:journals/bjet/Meij19}. 

\textbf{Disciplines:} Most of the experiments were performed using learning videos in STEM disciplines (around 80\%)~\cite{DBLP:journals/ce/BeegeNSNSWMR20, DBLP:journals/chb/MerktBFS18, DBLP:journals/chb/BeegeNSR19, DBLP:conf/educon/KleftodimosE18, DBLP:conf/hci/LeisnerZRC20, DBLP:conf/hicss/KellerLFL19, DBLP:conf/icls/DingASBC18, DBLP:journals/tkl/SchroederCC20, DBLP:journals/jche/MoonR21, DBLP:journals/ijcses/SozeriK21, DBLP:conf/ilrn/Munoz-CarpioCB20, DBLP:journals/ijim/Al-KhateebA20, DBLP:journals/mta/Perez-NavarroGC21, DBLP:conf/cscwd/LuLW21, DBLP:conf/collabtech/NugrahaWZHI20, DBLP:journals/bjet/Garrett21, DBLP:conf/chi/LeeM20, DBLP:conf/lak/LangCMP20, chen2021exploring, DBLP:conf/lak/SrivastavaNLVE020, DBLP:journals/chb/HorovitzM21, DBLP:conf/ecis/WinklerHFSSL20, DBLP:journals/ce/PiZZXYH19, DBLP:conf/chi/SrivastavaVLEB19, DBLP:journals/chb/StullFM18, DBLP:journals/jcal/WangPH19, DBLP:journals/jcal/PiHY17, DBLP:journals/bjet/PiHY17, DBLP:conf/sigcse/SchreiberD17, DBLP:journals/chb/LiKBJ16, DBLP:conf/hci/Zhang21, DBLP:conf/icls/SharmaDGD16, DBLP:journals/zmp/ZeeAPSG17, DBLP:journals/sle/WachtlerHZE16, DBLP:journals/chb/CojeanJ17, DBLP:journals/ets/OzdemirIS16, DBLP:journals/chb/HoogerheideWNG18, DBLP:conf/amcis/Garrett18, DBLP:conf/iiaiaai/KuYC19, DBLP:journals/eait/PalaigeorgiouP19, DBLP:journals/ile/HewL20, DBLP:conf/hvei/GuervosRPMDG19, DBLP:journals/jcal/MeijM16, shelton2016exploring, DBLP:conf/sac/PimentelYMZ19, DBLP:conf/sigcse/WhitneyD19, ohashi2019comparison, DBLP:conf/icetc/CaoNW18, DBLP:journals/bjet/Meij19} and some studies experimented with videos in STEM and non-STEM disciplines (e.g.,~\cite{DBLP:journals/ce/BeegeNSNSWMR20}).
Experiments with STEM topics have primarily considered physics, mathematics, and computer science subjects. 
In general, we identified the following STEM disciplines:
\begin{inparaenum}[(1)]
    \item physics and mathematics (19\% of controlled experiments)~\cite{DBLP:journals/chb/MerktBFS18, DBLP:conf/hci/LeisnerZRC20, DBLP:journals/mta/Perez-NavarroGC21, DBLP:journals/chb/HoogerheideWNG18, DBLP:journals/eait/PalaigeorgiouP19, chen2021exploring, DBLP:journals/chb/HorovitzM21, DBLP:journals/sle/WachtlerHZE16, DBLP:conf/iiaiaai/KuYC19, DBLP:journals/ile/HewL20, DBLP:journals/ijim/Al-KhateebA20, DBLP:conf/lak/SrivastavaNLVE020, DBLP:conf/chi/SrivastavaVLEB19},
    \item computer science (19\%)~\cite{DBLP:journals/ijcses/SozeriK21, DBLP:conf/ilrn/Munoz-CarpioCB20, DBLP:conf/chi/LeeM20, DBLP:conf/lak/LangCMP20, DBLP:conf/ecis/WinklerHFSSL20, DBLP:conf/sigcse/SchreiberD17, DBLP:journals/chb/LiKBJ16, DBLP:conf/sac/PimentelYMZ19, ohashi2019comparison, DBLP:conf/sigcse/WhitneyD19, DBLP:journals/jcal/WangPH19, DBLP:conf/collabtech/NugrahaWZHI20, chen2021exploring},
    \item biology, chemistry, and medicine (14\%)~\cite{DBLP:conf/icls/DingASBC18, DBLP:conf/hci/Zhang21, DBLP:conf/lak/SrivastavaNLVE020, DBLP:conf/chi/SrivastavaVLEB19, DBLP:journals/ce/PiZZXYH19, DBLP:conf/cscwd/LuLW21, DBLP:journals/zmp/ZeeAPSG17, DBLP:journals/chb/StullFM18, DBLP:journals/chb/BeegeNSR19, DBLP:conf/hvei/GuervosRPMDG19},
    \item software use, digital media and communication (13\%)~\cite{DBLP:journals/bjet/Meij19, DBLP:journals/jcal/MeijM16, DBLP:journals/bjet/Garrett21, DBLP:conf/amcis/Garrett18, DBLP:conf/educon/KleftodimosE18, DBLP:conf/hicss/KellerLFL19, DBLP:journals/jcal/PiHY17, DBLP:journals/bjet/PiHY17, DBLP:journals/ets/OzdemirIS16}, and
    \item environmental science and meteorology  (11\%) 
   ~\cite{DBLP:journals/bjet/Meij19, DBLP:journals/jcal/MeijM16, DBLP:journals/bjet/Garrett21, DBLP:conf/amcis/Garrett18, DBLP:journals/tkl/SchroederCC20, DBLP:journals/ce/BeegeNSNSWMR20, DBLP:journals/jche/MoonR21, DBLP:conf/icls/SharmaDGD16, DBLP:conf/icetc/CaoNW18, DBLP:journals/chb/CojeanJ17, shelton2016exploring, chen2021exploring}
 \end{inparaenum}
 Experiments on non-STEM topics focused mainly on language education (6\%)~\cite{DBLP:journals/bjet/Chen20, DBLP:journals/ijicte/KuhailA20, DBLP:journals/access/DavilaXSG21, DBLP:journals/bjet/HuangHC20}, history and geography (6\%)~\cite{DBLP:journals/ce/AltenPJK20, DBLP:journals/chb/WangLHS20, DBLP:journals/access/DaherS21, DBLP:journals/ce/BeegeNSNSWMR20}, and psychology (4\%)~\cite{DBLP:journals/ijicte/CooksonKH20, DBLP:journals/tkl/MoosB16, chen2021exploring}.
        
\textbf{Production style:} Most of the experiments reported the production style of the videos 
(69\% of controlled experiments). 
The results showed that most of them used slide-based videos (39\% of the experiments)~\cite{DBLP:conf/hci/LeisnerZRC20, DBLP:journals/bjet/WangLCWS19, DBLP:conf/chi/LeeM20, DBLP:conf/lak/SrivastavaNLVE020, DBLP:journals/chb/HorovitzM21, DBLP:journals/ce/PiZZXYH19, DBLP:conf/chi/SrivastavaVLEB19, DBLP:journals/jcal/WangPH19, DBLP:journals/bjet/PiHY17, DBLP:conf/icls/SharmaDGD16, DBLP:journals/chb/HoogerheideWNG18, DBLP:conf/icchp/KikusawaOKR16, DBLP:journals/tkl/MoosB16, DBLP:journals/ile/HewL20, DBLP:conf/sac/PimentelYMZ19, ohashi2019comparison, DBLP:conf/icetc/CaoNW18}. 
The presence of an instructor was also very common among the experiments (36\%)~\cite{DBLP:journals/ce/BeegeNSNSWMR20, DBLP:journals/chb/BeegeNSR19, DBLP:conf/aied/StrancM19, DBLP:conf/icls/DingASBC18, DBLP:conf/collabtech/NugrahaWZHI20, DBLP:journals/bjet/WangLCWS19, DBLP:conf/lak/SrivastavaNLVE020, DBLP:journals/chb/HorovitzM21, DBLP:journals/ce/PiZZXYH19, DBLP:journals/chb/StullFM18, DBLP:journals/bjet/PiHY17, DBLP:journals/chb/HoogerheideWNG18, DBLP:conf/icchp/KikusawaOKR16, DBLP:journals/ile/HewL20, DBLP:conf/sac/PimentelYMZ19, DBLP:conf/icetc/CaoNW18}.
Less frequent production styles included handwriting (e.g., Khan-style, classroom-style videos)  (16\%),~\cite{DBLP:journals/mta/Perez-NavarroGC21, DBLP:conf/chi/LeeM20, DBLP:conf/lak/SrivastavaNLVE020, DBLP:journals/chb/StullFM18, DBLP:conf/sigcse/SchreiberD17, DBLP:conf/icls/SharmaDGD16, DBLP:journals/ile/HewL20}, talking-head (16\%)~\cite{DBLP:journals/mta/Perez-NavarroGC21, DBLP:journals/access/DavilaXSG21, DBLP:journals/jcal/PiHY17, DBLP:journals/bjet/PiHY17, DBLP:journals/chb/LiKBJ16, DBLP:journals/ile/HewL20, DBLP:conf/icetc/CaoNW18}, screencasts (14\%)~\cite{DBLP:conf/hicss/KellerLFL19, DBLP:journals/bjet/Garrett21, DBLP:journals/ets/OzdemirIS16, DBLP:conf/amcis/Garrett18, DBLP:journals/jcal/MeijM16, DBLP:journals/bjet/Meij19}, productions with virtual reality and 360-degree characteristics (14\%)~\cite{DBLP:journals/bjet/Chen20, DBLP:conf/ilrn/Munoz-CarpioCB20, DBLP:journals/ce/Araiza-AlbaKMSS21, DBLP:journals/access/DaherS21, DBLP:journals/bjet/HuangHC20,DBLP:conf/hvei/GuervosRPMDG19}, and 
animations in videos (content created entirely with drawings, illustrations, or computer-generated effects) (14\%)~\cite{DBLP:journals/tkl/SchroederCC20, DBLP:journals/jche/MoonR21, DBLP:journals/ijicte/CooksonKH20, DBLP:journals/ijim/Al-KhateebA20, DBLP:conf/lak/SrivastavaNLVE020, DBLP:conf/chi/SrivastavaVLEB19, DBLP:journals/chb/LiKBJ16}. 
Interestingly, we noticed that animations 
were considered more frequently in the last three years~\cite{DBLP:journals/jche/MoonR21, DBLP:journals/ijicte/CooksonKH20, DBLP:journals/ijim/Al-KhateebA20, DBLP:conf/lak/SrivastavaNLVE020, DBLP:journals/tkl/SchroederCC20, DBLP:conf/chi/SrivastavaVLEB19}.
Please note that several studies used more than one production style (e.g.,~\cite{DBLP:conf/lak/SrivastavaNLVE020, DBLP:journals/bjet/PiHY17, DBLP:journals/ile/HewL20}).

\textbf{Sample size:} The experiment with the least number of participants included four learners~\cite{DBLP:conf/icchp/KikusawaOKR16}, the one with most of participants included 398 learners~\cite{shelton2016exploring}, the average was 94 participants. 
There was 
one experiment with a very large sample size of more than 12,000 participants, which was performed on a MOOC platform~\cite{DBLP:conf/lak/ZeeDSGGSPA18}.  
    
\subsubsection{Video Characteristics in Experiments}    

\begin{figure}[h]
\centering
\includegraphics[height=6cm]{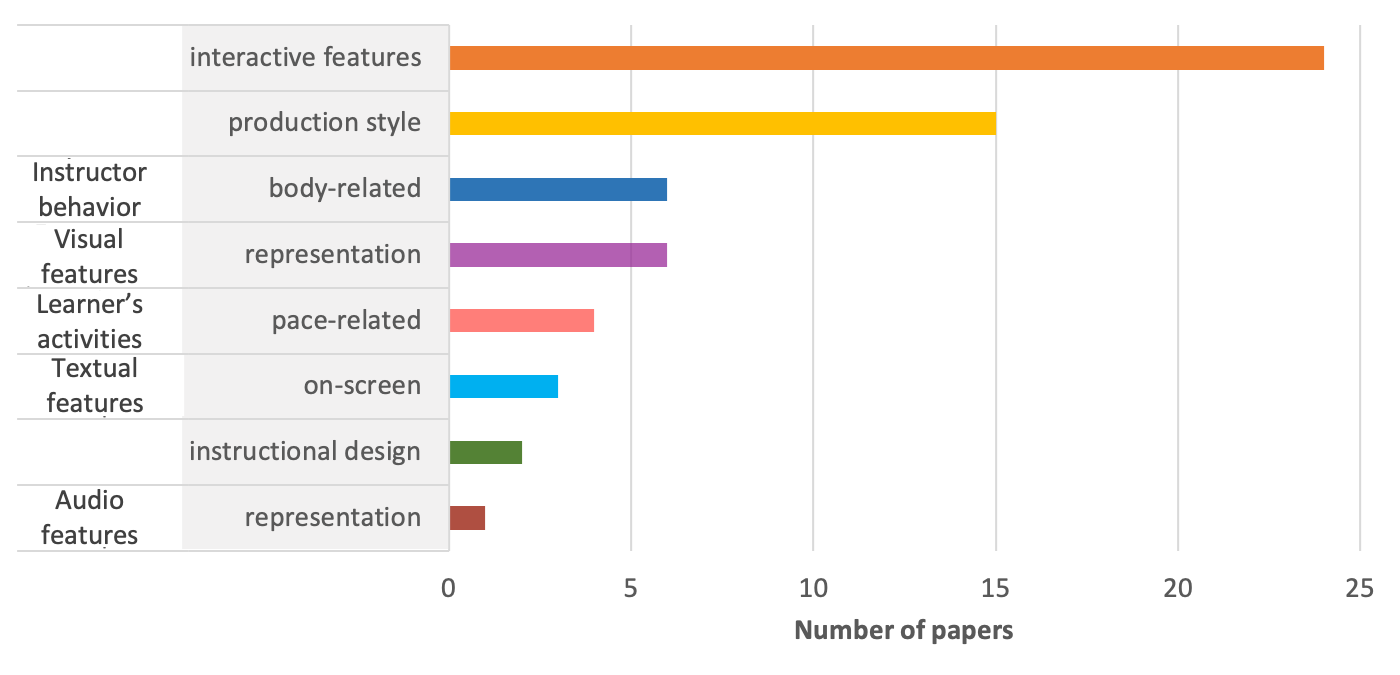}
\caption{The diagram shows the number of papers for each video feature type studied in controlled experiments. 
More than 80\% of these studies focused on 
interactive features, production style, instructor behavior, and visual features}. 
\label{features_experiments}
\end{figure}

Figure~\ref{features_experiments} summarizes and ranks the video characteristics studied in controlled experiments.
Specifically, we found four categories of characteristics that have been examined the most:
\begin{inparaenum}[(1)]
    \item interactive features (e.g., quizzes) (39\% of controlled experiments)~\cite{DBLP:conf/educon/KleftodimosE18, DBLP:journals/ce/AltenPJK20, DBLP:conf/hci/LeisnerZRC20, DBLP:conf/hicss/KellerLFL19,DBLP:conf/lak/ZeeDSGGSPA18, DBLP:journals/ijcses/SozeriK21, DBLP:conf/ectel/TaskinHHDM19,DBLP:conf/cscwd/LuLW21, DBLP:journals/ijicte/KuhailA20, chen2021exploring, DBLP:conf/ecis/WinklerHFSSL20, DBLP:journals/sle/WachtlerHZE16, DBLP:journals/chb/CojeanJ17, DBLP:conf/iiaiaai/KuYC19, DBLP:journals/eait/PalaigeorgiouP19, shelton2016exploring, DBLP:journals/tkl/MoosB16, DBLP:conf/sac/PimentelYMZ19, DBLP:journals/bjet/Chen20, DBLP:conf/ilrn/Munoz-CarpioCB20,DBLP:journals/ce/Araiza-AlbaKMSS21, DBLP:journals/access/DaherS21, DBLP:journals/bjet/HuangHC20, DBLP:conf/hvei/GuervosRPMDG19},
    \item production style (e.g., talking head) (25\%)~\cite{DBLP:conf/aied/StrancM19, DBLP:conf/icls/DingASBC18, DBLP:journals/ijicte/CooksonKH20, DBLP:journals/ijim/Al-KhateebA20, DBLP:journals/mta/Perez-NavarroGC21, DBLP:conf/collabtech/NugrahaWZHI20, DBLP:conf/chi/LeeM20, DBLP:conf/lak/SrivastavaNLVE020, DBLP:journals/access/DavilaXSG21, DBLP:conf/chi/SrivastavaVLEB19, DBLP:journals/chb/StullFM18, DBLP:journals/jcal/PiHY17, DBLP:journals/ile/HewL20, ohashi2019comparison, DBLP:conf/icetc/CaoNW18},
    \item instructor's behavior (e.g., gestures) (10\%)~\cite{DBLP:journals/ce/BeegeNSNSWMR20, DBLP:journals/chb/BeegeNSR19, DBLP:journals/bjet/WangLCWS19, DBLP:journals/chb/HorovitzM21, DBLP:journals/chb/LiKBJ16, DBLP:journals/chb/HoogerheideWNG18}, and
    \item visual features (10\%)~\cite{DBLP:journals/chb/WangLHS20, DBLP:journals/jche/MoonR21, DBLP:journals/ce/PiZZXYH19, DBLP:journals/jcal/WangPH19, DBLP:journals/bjet/PiHY17,  DBLP:conf/icls/SharmaDGD16}.
\end{inparaenum}

\textbf{Interactive features:}
These features allow the communication and interaction between the video and the learner.
The most used interactive features in experiments were multiple-choice quizzes and questions embedded in the video~\cite{DBLP:conf/educon/KleftodimosE18, DBLP:conf/hci/LeisnerZRC20, DBLP:journals/sle/WachtlerHZE16, DBLP:journals/eait/PalaigeorgiouP19, DBLP:conf/sac/PimentelYMZ19, DBLP:journals/ijcses/SozeriK21}, virtual reality and 360-degree environments~\cite{DBLP:journals/bjet/Chen20, DBLP:conf/ilrn/Munoz-CarpioCB20,DBLP:journals/ce/Araiza-AlbaKMSS21, DBLP:journals/access/DaherS21, DBLP:journals/bjet/HuangHC20, DBLP:conf/hvei/GuervosRPMDG19}, in-video annotations (feature that allow the learner to take notes while watching the video)~\cite{DBLP:conf/hci/LeisnerZRC20, DBLP:conf/ectel/TaskinHHDM19, DBLP:journals/ijicte/KuhailA20, chen2021exploring}, and navigational elements (e.g., table of content navigation, timelines with clickable points of interest)~\cite{DBLP:journals/ijicte/KuhailA20, DBLP:conf/sac/PimentelYMZ19, DBLP:journals/chb/CojeanJ17}. 
Examples of less researched interactive features were games (e.g., crossword puzzles)~\cite{DBLP:journals/chb/CojeanJ17}, human-like conversational agents~\cite{DBLP:conf/ecis/WinklerHFSSL20}, and 
self-regulated learning activities~\cite{DBLP:journals/ce/AltenPJK20, DBLP:journals/tkl/MoosB16}. 

\textbf{Production style:} 
In general, the production style refers to the scenery, the elements used to convey the information, and also the physical arrangements considered to record the video. 
Examples for production style settings are 
Khan-style video, classroom-style videos, with talking head, etc. 
Specifically, we use 
the typology of video production styles of Hansch et al.~\cite{hansch2015video}. 
We found that more than a third of the experiments that examined production styles 
addressed the 
instructor's presence~\cite{DBLP:conf/icetc/CaoNW18, DBLP:journals/ile/HewL20, DBLP:journals/jcal/PiHY17, DBLP:journals/access/DavilaXSG21, DBLP:journals/mta/Perez-NavarroGC21}. 
Another third of the experiments evaluated dialogue-style in contrast to the commonly used monologue-style~\cite{DBLP:conf/aied/StrancM19, DBLP:conf/icls/DingASBC18, DBLP:conf/collabtech/NugrahaWZHI20, DBLP:conf/chi/LeeM20, ohashi2019comparison}. 
The last third of the experiments in this category investigated animations~\cite{DBLP:journals/ijicte/CooksonKH20, DBLP:journals/ijim/Al-KhateebA20, DBLP:conf/lak/SrivastavaNLVE020, DBLP:conf/chi/SrivastavaVLEB19}.

\textbf{Visual and audio features:} This kind of experiments 
primarily addressed characteristics that attract the attention of the viewer, here called representation characteristics~\cite{DBLP:journals/chb/WangLHS20, DBLP:journals/jche/MoonR21, DBLP:journals/ce/PiZZXYH19, DBLP:journals/jcal/WangPH19, DBLP:journals/bjet/PiHY17, DBLP:conf/icls/SharmaDGD16}. 
Experiments on representation characteristics have clearly focused 
on content-related cues that direct attention (e.g., arrows, moving elements, color cues, gaze overlay)~\cite{DBLP:journals/chb/WangLHS20, DBLP:journals/jche/MoonR21, DBLP:journals/ce/PiZZXYH19, DBLP:conf/icls/SharmaDGD16} and instructor-related elements that direct attention (e.g., pointing gestures, spoken words, directed eye gaze)~\cite{DBLP:journals/chb/WangLHS20, DBLP:journals/jche/MoonR21, DBLP:journals/ce/PiZZXYH19, DBLP:journals/jcal/WangPH19, DBLP:journals/bjet/PiHY17, DBLP:conf/icls/SharmaDGD16}.
Other experiments studied the impact of the complexity of visual content~\cite{DBLP:conf/hvei/GuervosRPMDG19} and the gender appearance of the instructor~\cite{DBLP:journals/chb/HoogerheideWNG18}.

\textbf{Instructor behavior characteristics:} In controlled experiments, the behavior of the instructors has been studied principally with respect to their gestures~\cite{DBLP:journals/ce/BeegeNSNSWMR20, DBLP:journals/chb/BeegeNSR19, DBLP:journals/bjet/WangLCWS19, DBLP:journals/chb/HorovitzM21, DBLP:journals/chb/LiKBJ16}. 
Less studied characteristics were related to the speech (e.g., tone of voice~\cite{DBLP:journals/chb/HorovitzM21}), other visual aspects such as pose (e.g., frontal, lateral), and dressing style~\cite{DBLP:journals/chb/BeegeNSR19}.

\textbf{Other characteristics:} 
Some articles studied the pace at which videos are viewed 
\cite{DBLP:journals/chb/MerktBFS18, DBLP:journals/tkl/SchroederCC20, DBLP:journals/bjet/Garrett21, DBLP:conf/lak/LangCMP20, DBLP:conf/amcis/Garrett18}. 
Among the less popular researched topics, we found experiments with captioned and non-captioned videos~\cite{DBLP:journals/ets/OzdemirIS16, DBLP:conf/sigcse/WhitneyD19}, incorporating music~\cite{DBLP:conf/sigcse/SchreiberD17}, summaries at the end of video~\cite{DBLP:journals/jcal/MeijM16}, the inclusion of sign language~\cite{DBLP:conf/icchp/KikusawaOKR16}, and experiments that test Multimedia Learning Theory such as the effect of pre-training material and spatial contiguity between captions and visual elements~\cite{DBLP:conf/hci/Zhang21, DBLP:journals/bjet/Meij19}.

\subsubsection{Dependent Variables}

In controlled experiments, the dependent variable measured the impact of specific video characteristics on different learning aspects. 
We have reviewed all the dependent variables and categorized them into different groups according to their purpose.
In this way, we have obtained the following categories that are related to:
\begin{inparaenum}[(a)]
    \item self-perceived effectiveness (e.g., self-efficacy, instructional efficiency)~\cite{DBLP:journals/ile/HewL20, DBLP:journals/chb/HoogerheideWNG18, DBLP:conf/hicss/KellerLFL19, DBLP:journals/chb/CojeanJ17,DBLP:journals/jcal/MeijM16, DBLP:conf/sac/PimentelYMZ19, DBLP:journals/tkl/SchroederCC20, DBLP:journals/tkl/MoosB16, DBLP:conf/chi/LeeM20, DBLP:journals/bjet/Meij19, DBLP:journals/bjet/HuangHC20, DBLP:journals/ijcses/SozeriK21}, 
    \item difficulty (e.g., cognitive load, time effort)~\cite{DBLP:conf/icetc/CaoNW18, DBLP:conf/hicss/KellerLFL19, DBLP:journals/chb/HoogerheideWNG18, DBLP:journals/chb/CojeanJ17, DBLP:conf/sac/PimentelYMZ19, DBLP:journals/ce/BeegeNSNSWMR20, DBLP:conf/cscwd/LuLW21, DBLP:conf/lak/SrivastavaNLVE020, DBLP:conf/chi/SrivastavaVLEB19, DBLP:journals/zmp/ZeeAPSG17, DBLP:journals/tkl/MoosB16, DBLP:journals/chb/MerktBFS18, DBLP:journals/chb/BeegeNSR19, DBLP:journals/bjet/Garrett21, DBLP:conf/chi/LeeM20, chen2021exploring, DBLP:journals/access/DavilaXSG21, DBLP:journals/bjet/PiHY17, DBLP:journals/jche/MoonR21, DBLP:journals/bjet/HuangHC20, DBLP:journals/jcal/PiHY17},
    \item general learning experience (e.g., lecture experience, learning experience)~\cite{DBLP:journals/chb/StullFM18, DBLP:journals/chb/LiKBJ16, DBLP:journals/eait/PalaigeorgiouP19, shelton2016exploring, DBLP:conf/educon/KleftodimosE18, DBLP:journals/ce/AltenPJK20, DBLP:journals/ijicte/CooksonKH20, DBLP:journals/mta/Perez-NavarroGC21, DBLP:conf/cscwd/LuLW21, DBLP:journals/bjet/WangLCWS19, DBLP:journals/bjet/HuangHC20, DBLP:journals/jcal/PiHY17, DBLP:conf/icetc/CaoNW18, DBLP:journals/bjet/Chen20},
    \item emotions and feelings of the viewer (e.g., satisfaction, enjoyment)~\cite{DBLP:journals/bjet/WangLCWS19, DBLP:journals/chb/HorovitzM21, DBLP:journals/ce/Araiza-AlbaKMSS21, DBLP:journals/chb/HoogerheideWNG18, DBLP:conf/lak/SrivastavaNLVE020, DBLP:conf/chi/SrivastavaVLEB19, DBLP:conf/sigcse/SchreiberD17, DBLP:conf/aied/StrancM19, DBLP:journals/chb/CojeanJ17, DBLP:conf/hvei/GuervosRPMDG19},
    \item instructor (e.g., enthusiasm, presentation skills)~\cite{DBLP:journals/chb/LiKBJ16, DBLP:journals/chb/HoogerheideWNG18, DBLP:journals/ijim/Al-KhateebA20, DBLP:journals/ile/HewL20, DBLP:journals/chb/CojeanJ17, DBLP:conf/chi/LeeM20, DBLP:journals/access/DavilaXSG21, DBLP:journals/jcal/PiHY17, DBLP:journals/jcal/WangPH19, DBLP:conf/icetc/CaoNW18}, 
    \item video and its content (e.g., quality, usability)~\cite{DBLP:conf/hvei/GuervosRPMDG19, DBLP:conf/icchp/KikusawaOKR16, DBLP:journals/eait/PalaigeorgiouP19, ohashi2019comparison, DBLP:conf/hicss/KellerLFL19, DBLP:journals/access/DaherS21, shelton2016exploring, DBLP:journals/jcal/MeijM16} 
    \item engagement with the video (e.g., motivation, engagement) ~\cite{DBLP:conf/ilrn/Munoz-CarpioCB20, DBLP:journals/ijicte/CooksonKH20, DBLP:conf/ecis/WinklerHFSSL20, shelton2016exploring, DBLP:conf/hci/LeisnerZRC20, DBLP:journals/ce/Araiza-AlbaKMSS21,DBLP:conf/chi/LeeM20, DBLP:conf/chi/SrivastavaVLEB19, DBLP:journals/jcal/WangPH19, DBLP:conf/cscwd/LuLW21, DBLP:conf/icls/SharmaDGD16, DBLP:journals/bjet/Chen20, DBLP:journals/bjet/HuangHC20, DBLP:journals/chb/HorovitzM21, DBLP:journals/ets/OzdemirIS16, DBLP:conf/hicss/KellerLFL19},
    \item learning performance (e.g., learning quality, learning outcome) ~\cite{DBLP:conf/hicss/KellerLFL19, DBLP:journals/access/DaherS21, DBLP:journals/chb/CojeanJ17, DBLP:journals/chb/MerktBFS18, DBLP:journals/chb/BeegeNSR19, DBLP:conf/educon/KleftodimosE18, DBLP:journals/ce/AltenPJK20, DBLP:conf/hci/LeisnerZRC20, DBLP:journals/bjet/Chen20, DBLP:conf/icls/DingASBC18, DBLP:conf/lak/ZeeDSGGSPA18, DBLP:journals/chb/WangLHS20, DBLP:journals/tkl/SchroederCC20, DBLP:journals/jche/MoonR21, DBLP:journals/ijcses/SozeriK21, DBLP:journals/ijicte/CooksonKH20, DBLP:journals/ijim/Al-KhateebA20, DBLP:journals/ce/Araiza-AlbaKMSS21, DBLP:conf/ectel/TaskinHHDM19, DBLP:conf/cscwd/LuLW21, DBLP:conf/collabtech/NugrahaWZHI20, DBLP:journals/bjet/WangLCWS19, DBLP:journals/ijicte/KuhailA20, DBLP:journals/bjet/HuangHC20, DBLP:journals/bjet/Garrett21, DBLP:conf/chi/LeeM20, DBLP:conf/lak/LangCMP20, DBLP:conf/lak/SrivastavaNLVE020, DBLP:journals/chb/HorovitzM21, DBLP:journals/access/DavilaXSG21, DBLP:journals/ce/PiZZXYH19, DBLP:journals/chb/StullFM18, DBLP:journals/jcal/WangPH19, DBLP:conf/iiaiaai/Lin0L18, DBLP:journals/jcal/PiHY17, DBLP:journals/bjet/PiHY17, DBLP:conf/sigcse/SchreiberD17, DBLP:journals/chb/LiKBJ16, DBLP:conf/hci/Zhang21, DBLP:conf/icls/SharmaDGD16, DBLP:journals/zmp/ZeeAPSG17, DBLP:journals/sle/WachtlerHZE16, DBLP:journals/chb/CojeanJ17, DBLP:journals/ets/OzdemirIS16, DBLP:journals/chb/HoogerheideWNG18, DBLP:conf/amcis/Garrett18, DBLP:conf/iiaiaai/KuYC19, DBLP:journals/eait/PalaigeorgiouP19, DBLP:journals/tkl/MoosB16, DBLP:journals/ile/HewL20, DBLP:journals/jcal/MeijM16, shelton2016exploring, DBLP:conf/sac/PimentelYMZ19, DBLP:conf/sigcse/WhitneyD19, llanda2019video, DBLP:conf/icetc/CaoNW18, DBLP:journals/bjet/Meij19}, and
    \item other (e.g., cybersickness, perceived similarity of the learner to the instructor)~\cite{DBLP:journals/bjet/HuangHC20, DBLP:conf/hvei/GuervosRPMDG19, DBLP:journals/chb/CojeanJ17, DBLP:journals/chb/BeegeNSR19, ohashi2019comparison, DBLP:journals/chb/HoogerheideWNG18, DBLP:journals/ile/HewL20}.
\end{inparaenum}

Specifically, within these categories, the evaluated variable in almost every experiment was the learning outcome (86\% of controlled experiments), followed by cognitive load (20\%), satisfaction (19\%), mental effort (11\%), social presence (learners' perception of their own social presence) (11\%), motivation (9\%), and self-efficacy (9\%).
Please note that several studies measured the effects on more than one dependent variable.

The learning outcome has been studied under different names such as learning gain, academic achievement, etc.
Lately, this metric has been measured not only by testing how much the learner can remember (\emph{recall}) but also to what extent the learner can apply the gained knowledge to new problems (\emph{transfer}). 
Our results showed that transfer tests have been more frequently used in the last four years than in the previous years~\cite{DBLP:journals/chb/CojeanJ17, DBLP:journals/chb/MerktBFS18, DBLP:journals/chb/BeegeNSR19, DBLP:journals/ce/AltenPJK20, DBLP:conf/icls/DingASBC18, DBLP:journals/chb/WangLHS20, DBLP:journals/tkl/SchroederCC20, DBLP:conf/cscwd/LuLW21, DBLP:journals/ijicte/KuhailA20, DBLP:journals/bjet/Garrett21, DBLP:journals/ce/PiZZXYH19, DBLP:conf/hci/Zhang21, DBLP:conf/amcis/Garrett18, DBLP:conf/iiaiaai/KuYC19, DBLP:journals/ile/HewL20}. 
However, we found that most of the articles 
did not report explicitly 
if the type of test was recall or transfer,
only 38\% of experiments reported it~\cite{DBLP:journals/chb/CojeanJ17, DBLP:journals/chb/MerktBFS18, DBLP:journals/chb/BeegeNSR19, DBLP:journals/ce/AltenPJK20, DBLP:conf/icls/DingASBC18, DBLP:journals/chb/WangLHS20, DBLP:journals/tkl/SchroederCC20, DBLP:conf/cscwd/LuLW21, DBLP:journals/ijicte/KuhailA20, DBLP:journals/ce/PiZZXYH19, DBLP:conf/hci/Zhang21, DBLP:conf/iiaiaai/KuYC19, DBLP:journals/ile/HewL20, DBLP:conf/educon/KleftodimosE18, DBLP:journals/ce/Araiza-AlbaKMSS21, DBLP:conf/collabtech/NugrahaWZHI20, DBLP:journals/access/DavilaXSG21, DBLP:journals/chb/LiKBJ16, DBLP:conf/sigcse/WhitneyD19, DBLP:journals/bjet/Garrett21, DBLP:conf/amcis/Garrett18}. 

In addition, we found that dependent variables were derived from tracking facial expressions~\cite{DBLP:conf/chi/SrivastavaVLEB19}, neurophysiological aspects (e.g., galvanic skin)~\cite{DBLP:conf/ecis/WinklerHFSSL20}, thermal aspects~\cite{DBLP:conf/chi/SrivastavaVLEB19, DBLP:conf/lak/SrivastavaNLVE020}, log data~\cite{DBLP:conf/sac/PimentelYMZ19, DBLP:conf/lak/LangCMP20}, and eye gaze~\cite{DBLP:conf/collabtech/NugrahaWZHI20, DBLP:conf/chi/LeeM20, DBLP:journals/chb/StullFM18, DBLP:journals/jcal/WangPH19, DBLP:conf/icchp/KikusawaOKR16, DBLP:journals/ce/PiZZXYH19, DBLP:conf/hci/Zhang21, DBLP:conf/icls/SharmaDGD16, DBLP:journals/jche/MoonR21, DBLP:journals/bjet/PiHY17, DBLP:conf/chi/SrivastavaVLEB19}. 
The last one was the most used technique, from which popular metrics such as dwell time have been derived to measure, for example, attention.

\subsection{Data Analysis Studies}
\label{subsec: data analysis studies}

The majority of data analysis studies focused 
on the understanding of patterns in learning behavior. 
Furthermore, around half of the data analysis papers have also tried to understand the impact of those patterns on a specific variable that measures learning effectiveness~\cite{DBLP:conf/helmeto/EradzeDFP20, DBLP:journals/bjet/StohrSMNM19, DBLP:conf/amia/WynnWLSBWC19, DBLP:journals/caee/GranjoR18, DBLP:journals/jche/CostleyFLB21, DBLP:conf/sigcse/AngraveZHM20, DBLP:journals/ijhci/Li19, DBLP:conf/educon/RodriguezNFPM18, DBLP:journals/itse/CostleyL17, DBLP:journals/tsp/BrintonBCP16, DBLP:conf/lats/Kovacs16, DBLP:journals/oir/LeeOGK17, DBLP:journals/i-jep/HildebrandA18, DBLP:journals/ce/Shoufan19, DBLP:conf/icse/GalsterMG18, DBLP:conf/lats/ThorntonRW17, DBLP:conf/lak/AtapattuF17}. 
In the same way as in the controlled experiment papers, the impact of video characteristics has been studied mostly on the learning outcome~\cite{DBLP:journals/itse/CostleyL17, DBLP:journals/bjet/StohrSMNM19, DBLP:conf/amia/WynnWLSBWC19, DBLP:conf/sigcse/AngraveZHM20, DBLP:journals/ce/Shoufan19, DBLP:conf/icse/GalsterMG18, DBLP:conf/educon/RodriguezNFPM18, DBLP:journals/tsp/BrintonBCP16, DBLP:conf/lats/Kovacs16, DBLP:journals/i-jep/HildebrandA18, DBLP:conf/lats/ThorntonRW17, DBLP:journals/ijhci/Li19} and 
on engagement~\cite{DBLP:journals/caee/GranjoR18, DBLP:journals/ijhci/Li19, DBLP:journals/itse/CostleyL17, DBLP:conf/icse/GalsterMG18, DBLP:conf/lats/ThorntonRW17, DBLP:conf/lak/AtapattuF17, DBLP:journals/i-jep/HildebrandA18}.

\begin{figure}[h]
\centering
\includegraphics[height=7cm]{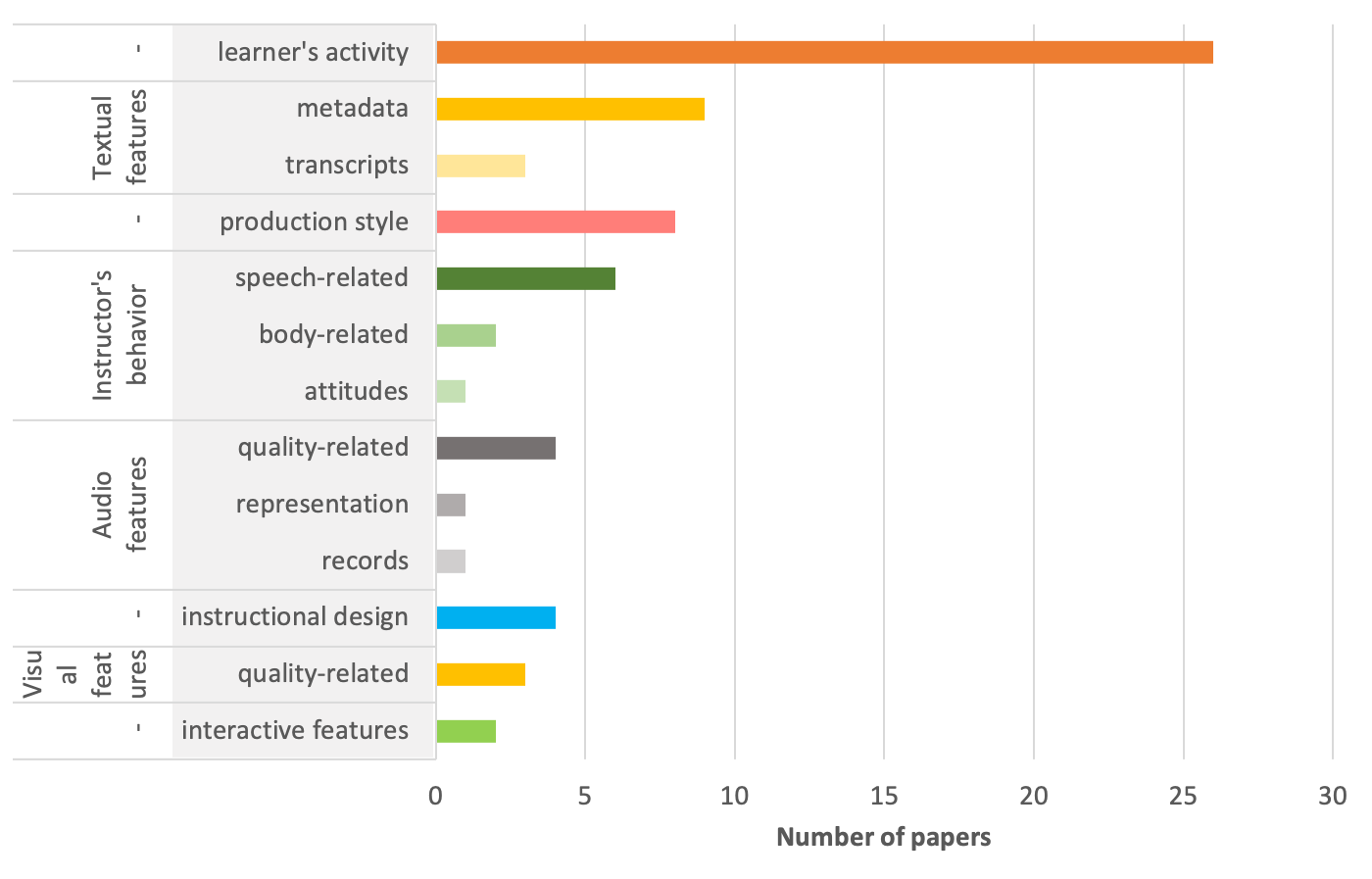}
\caption{The diagram displays the number of papers with regard to video characteristics explored in their data analysis studies. These studies usually made 
use of log data of learner's activities.}
\label{features_danalysis}
\end{figure}

Figure~\ref{features_danalysis} sorts the characteristics studied in data analysis articles according to their frequency.
We found that data analysis studies usually made 
use of 
learner's activity characteristics by taking advantage of system logs (e.g., play, pause)~\cite{DBLP:conf/helmeto/EradzeDFP20, DBLP:journals/ce/SaurabhG19, DBLP:journals/chb/Shoufan19, DBLP:journals/bjet/StohrSMNM19, DBLP:conf/amia/WynnWLSBWC19, DBLP:journals/jche/CostleyFLB21, DBLP:conf/sigcse/AngraveZHM20, DBLP:journals/ijhci/Li19, DBLP:conf/educon/RodriguezNFPM18, DBLP:journals/itse/CostleyL17, DBLP:journals/bjet/GiannakosJK16, DBLP:journals/nca/HuZGW20, DBLP:journals/aiedu/UchidiunoKHYO18, DBLP:conf/lats/Kovacs16, DBLP:conf/lats/SluisGZ16, DBLP:journals/ijmlo/LeiYLTYL19, DBLP:journals/ce/Shoufan19, DBLP:journals/tele/Meseguer-Martinez19, DBLP:journals/elearn/YanB18, DBLP:conf/lak/AtapattuF17, DBLP:journals/oir/LeeOGK17, DBLP:conf/lak/LeiGHOQKYL17, DBLP:conf/chiir/DodsonRFYHF18, DBLP:conf/cchi/PatelGZ18}. 
Other characteristics often analyzed were metadata 
and video transcripts~\cite{DBLP:conf/helmeto/EradzeDFP20, DBLP:journals/ce/SaurabhG19, DBLP:journals/chb/Shoufan19, DBLP:journals/ijcallt/Ding18, DBLP:journals/corr/abs-2005-13876, DBLP:journals/oir/LeeOGK17, silva2017instructional, DBLP:journals/i-jep/HildebrandA18, DBLP:conf/ectel/SjodenDM18, DBLP:conf/lats/SluisGZ16, DBLP:journals/tele/Meseguer-Martinez19, DBLP:conf/icse/GalsterMG18, DBLP:conf/lak/AtapattuF17}. 
Moreover, data analysis studies have inspected production style characteristics~\cite{DBLP:journals/ijcallt/Ding18, DBLP:journals/itse/CostleyL17, DBLP:journals/i-jep/HildebrandA18, rahim2019video, DBLP:conf/lats/ThorntonRW17, DBLP:journals/chb/Shoufan19, DBLP:journals/ijmlo/LeiYLTYL19, DBLP:conf/sigite/ShoufanM17},
audio features~\cite{DBLP:journals/corr/abs-2005-13876, DBLP:journals/chb/Shoufan19, DBLP:journals/ijcallt/Ding18, DBLP:conf/sigite/ShoufanM17} and visual features~\cite{DBLP:journals/chb/Shoufan19, DBLP:journals/ijcallt/Ding18, DBLP:conf/sigite/ShoufanM17}, 
instructor's behavior characteristics~\cite{DBLP:journals/ijcallt/Ding18, DBLP:conf/ecce/TianB16, DBLP:journals/chb/Shoufan19, DBLP:conf/lats/SluisGZ16, DBLP:conf/lak/AtapattuF17, DBLP:conf/sigite/ShoufanM17, DBLP:conf/cchi/PatelGZ18},  
design principles and guidelines~\cite{DBLP:conf/helmeto/EradzeDFP20, DBLP:journals/chb/Shoufan19, DBLP:conf/lats/OuGJH16, DBLP:journals/ijmlo/LeiYLTYL19}, and 
interactive features~\cite{DBLP:journals/ijcallt/Ding18, DBLP:journals/caee/GranjoR18}. 
The corresponding findings are presented in the next section (Section~\ref{sec:findings}).

\subsection{Design Guidelines}

\textbf{General guidelines:} Wijnker et al.~\cite{DBLP:journals/bjet/WijnkerBGD19} developed a framework that applies a film theory perspective to learning videos.
The authors provided guidance to analyze and categorize these videos according to their design, aim and type in film theory~\cite{bordwell1993film}.
Voronkin~\cite{DBLP:conf/icteri/Voronkin19} proposed universal guidelines focusing on methodological, psychological, didactic, ergonomic, and technical aspects. 
Lange and Costley~\cite{lange2020improving} considered media delivery problems and addressed them by proposing guidelines that aim at improving the pace, intelligibility, quality, media diversity, and congruence.
The authors grounded their guidelines in the limited capacity theory in multimedia learning~\cite{mayer2014multimedia},  
generative theory~\cite{mayer2014multimedia}, and 
extraneous processing research~\cite{leppink2013development}. 
Dodson et al.~\cite{DBLP:conf/lats/DodsonRFYHF18} introduced a framework of active viewing that categorized learner's watching behavior. 
The authors stated that VBL can be designed in a way to promote this behavior. 

\textbf{Guidelines for specific disciplines}: Minnes et al.~\cite{DBLP:conf/sigcse/MinnesAGF19} proposed design guidelines for producing short learning podcast videos for computer science courses and emphasized the use of representation characteristics, especially call-out boxes, zooming, and animations.
Hodges~\cite{hodges2018ensuring} also collected several recommendations for creating videos in CS. 
The authors emphasized the need of captions, descriptions of animations as they occur, zooming in relevant code, and for video systems that allow learners to pause, seek, and speed up or slow down the video.
Stephenson~\cite{DBLP:conf/sigcse/Stephenson19} concentrated on the development of programming videos following a specific content structure.
Lei et al.~\cite{DBLP:conf/tale/LeiYKLA16} discussed an approach to produce videos for Humanities disciplines. They suggested to favor short videos, use animations, highlight key concepts and terminologies, include an introduction and conclusions, employ consistent topography and graphic styles, and to add music in specific parts.
Ge and Li~\cite{DBLP:conf/iceit/GeL20} identified beneficial characteristics to learn English as a foreign language. 
They recommended to take care of the speech rate, sentences and vocabulary complexity, use of closely related examples, use of realistic visuals when higher levels of context are demanded, include the instructor, and use keyword subtitles instead of complete subtitles.

\textbf{Guidelines for videos with interactive features} 
Weinert et al.~\cite{DBLP:conf/icis/WeinertGBB20} introduced reference guidelines to design 
Interaction Theory~\cite{moore1989three}, Cognitive Load Theory~\cite{sweller1994cognitive}, and own research. 
The authors recommended to include elements for interacting not only with the content but also with other learners and the instructor.
They also discussed recommendations for the best positions to place specific types of interactive features.
Hung et al.~\cite{DBLP:journals/ce/HungKC18} proposed guidelines to design interactive tasks for learning video content based on literature on embodied cognition. 
The authors propose that the video should include engaging, prompting, experiencing, facilitating, demonstrating, and questioning tasks.

\section{Impact of video characteristics on the learning effectiveness}
\label{sec:findings}

In this section, we contrast, analyze, and summarize the findings of studies from controlled experiments and data analysis research. These are organized according to the different categories and subcategories of video characteristics.

\subsection{Instructor Behavior Characteristics}
    
\textbf{Gestures:} Study results showed that gestures do not have a significant impact on the learning outcome~\cite{DBLP:journals/ce/BeegeNSNSWMR20, DBLP:journals/bjet/WangLCWS19, DBLP:journals/chb/HorovitzM21}. 
One study even found that gestures can cause negative comments~\cite{DBLP:journals/ijcallt/Ding18}. 
However, learners still care about gestures, facial expressions, and body language~\cite{DBLP:journals/chb/Shoufan19}.
Among the benefits, gestures were found to reduce mental effort~\cite{DBLP:journals/ce/BeegeNSNSWMR20}, increase presence~\cite{DBLP:journals/ce/BeegeNSNSWMR20, DBLP:journals/chb/LiKBJ16} and satisfaction~\cite{DBLP:journals/bjet/WangLCWS19}.
Moreover, studies have obtained a positive effect on delayed learning outcome tests and a higher impact of human gestures compared to gestures of instructors as animations (human-like artificial agents)~\cite{DBLP:journals/bjet/WangLCWS19, DBLP:journals/chb/HorovitzM21}. 

\textbf{Speech:} Data analysis studies revealed that characteristics related to the speech such as speaking rate, lexical diversity, etc., affected how learners interacted with the video~\cite{DBLP:conf/lak/AtapattuF17}. 
Speech characteristics were also found to correlate with sound characteristics~\cite{DBLP:journals/corr/abs-2005-13876}.
Further findings showed that speech characteristics had an impact on dwell time~\cite{DBLP:conf/lats/SluisGZ16} and the level of understanding~\cite{DBLP:journals/chb/Shoufan19}. 
For example, whether the instructor's language was native or not significantly affected the level of understanding in the learner, and the number of likes in the video~\cite{DBLP:journals/chb/Shoufan19, DBLP:conf/sigite/ShoufanM17}.
Also, pronunciation and intonation was found to affect the learner's attention.
Based on these discoveries, it has been recommended to reduce lexical diversity, long sentences, and speech rate~\cite{DBLP:conf/lak/AtapattuF17}. 

\textbf{Other:} Other characteristics related to instructor behavior which the learners cared about and had an impact were the appearance, clothes, and enthusiasm~\cite{DBLP:journals/chb/Shoufan19, DBLP:journals/ijcallt/Ding18}. 
Interestingly, the gender did not have a significant impact~\cite{DBLP:journals/chb/Shoufan19}, it did not affect the learning outcome, the perception of self-efficacy~\cite{DBLP:journals/chb/HoogerheideWNG18}, and learners liking or disliking a video~\cite{DBLP:conf/sigite/ShoufanM17}. 
Yet, it was observed that learners perceived themselves more similar to the instructor when the gender was the same~\cite{DBLP:journals/chb/HoogerheideWNG18}. 
With regard to instructors as animations, these were considered interesting and attractive~\cite{DBLP:journals/ijcallt/Ding18}.

\subsection{Interactive Features}

\textbf{Quizzes and questions}: Most of the studies that experimented with quizzes evidenced that the learning outcome was positively affected~\cite{DBLP:conf/educon/KleftodimosE18, DBLP:conf/hci/LeisnerZRC20, DBLP:journals/sle/WachtlerHZE16, DBLP:journals/eait/PalaigeorgiouP19, DBLP:conf/sac/PimentelYMZ19}.
For Palaigeorgiou and Papadopoulou~\cite{DBLP:journals/eait/PalaigeorgiouP19}, the positive effect was found when quizzes were included together with other interactive features.
For Wachtler et al.~\cite{DBLP:journals/sle/WachtlerHZE16}, the positive effect was found when quizzes were placed at the right time. 

Furthermore, positive impacts were found on other aspects such as satisfaction~\cite{DBLP:conf/educon/KleftodimosE18, DBLP:journals/eait/PalaigeorgiouP19} and engagement~\cite{DBLP:conf/sac/PimentelYMZ19}. 
Data analysis studies have also confirmed these findings. 
Kovacs~\cite{DBLP:conf/lats/Kovacs16} found that most viewers engaged in quizzes, which seemed to reduce dropout.
On the contrary, Ding~\cite{DBLP:journals/ijcallt/Ding18} showed that quizzes 
improved understanding of the content. But they did not draw the learner's attention in cases where the quizzes were complex and at the end of the video.
        
\textbf{Virtual reality and 360-degree characteristics:} 
The studies in this categories commonly reported an improved learning outcome, however, not to a significant degree~\cite{DBLP:journals/bjet/Chen20, DBLP:journals/ce/Araiza-AlbaKMSS21, DBLP:journals/access/DaherS21, DBLP:journals/bjet/HuangHC20}. 
Huang et al.~\cite{DBLP:journals/bjet/HuangHC20} found positive effects only in certain parts of the post test. 
Although statistically insignificant, Araiza-Alba et al.~\cite{DBLP:journals/ce/Araiza-AlbaKMSS21} reported short- and long-term improvements in learning outcome.
Studies also showed differing impact of virtual reality and 360-degree environments on learner motivation. 
While Huang et al.~\cite{DBLP:journals/bjet/HuangHC20} found no positive effect on the motivation, Muñoz-Carpio et al.~\cite{DBLP:conf/ilrn/Munoz-CarpioCB20} and Chen~\cite{DBLP:journals/bjet/Chen20} found that viewers felt indeed more motivated.
Regarding satisfaction, several findings confirmed an increase~\cite{DBLP:journals/bjet/Chen20, DBLP:conf/ilrn/Munoz-CarpioCB20, DBLP:journals/bjet/HuangHC20}. 
Furthermore, Huang et al.~\cite{DBLP:journals/bjet/HuangHC20} also found a positive impact on self-efficacy.
Our analysis over time shows that these kind of characteristics have been investigated more frequently in the last years~\cite{DBLP:journals/bjet/Chen20, DBLP:conf/ilrn/Munoz-CarpioCB20, DBLP:journals/bjet/HuangHC20, DBLP:journals/ce/Araiza-AlbaKMSS21, DBLP:journals/access/DaherS21}.
Naturally, experiments in this category evaluated the impact in other non-common dependent variables such as sense of "being there"~\cite{DBLP:journals/ce/Araiza-AlbaKMSS21, DBLP:conf/hvei/GuervosRPMDG19} and cybersickness~\cite{DBLP:conf/hvei/GuervosRPMDG19}.
Interestingly, most of these experiments were conducted with non-STEM videos~\cite{DBLP:journals/bjet/Chen20, DBLP:journals/ce/Araiza-AlbaKMSS21, DBLP:journals/access/DaherS21, DBLP:journals/bjet/HuangHC20}.

\textbf{Other:} Most of the studies that experimented with in-video annotations found an increase in the learning outcome~\cite{DBLP:conf/hci/LeisnerZRC20,DBLP:conf/ectel/TaskinHHDM19,DBLP:journals/ijicte/KuhailA20}, whereas neutral effects were found for cognitive load~\cite{chen2021exploring}.
Regarding navigational features, experiments using table content navigation~\cite{DBLP:journals/ijicte/KuhailA20, DBLP:journals/chb/CojeanJ17} and timeslines with clickable points~\cite{DBLP:conf/sac/PimentelYMZ19, DBLP:journals/chb/CojeanJ17} found a positive effect on the learning outcome~\cite{DBLP:journals/ijicte/KuhailA20, DBLP:conf/sac/PimentelYMZ19, DBLP:journals/chb/CojeanJ17}. 
However, this was not always significant~\cite{DBLP:conf/sac/PimentelYMZ19}, 
or depended 
on the combination with other features to be more beneficial such as table content navigation together with timelines with markers~\cite{DBLP:journals/chb/CojeanJ17}.
According to Granjo and Rasteiro~\cite{DBLP:journals/caee/GranjoR18}, 
videos with interactive features needed to be used with other pedagogical techniques to improve the learning outcome.

\subsection{Visual and Audio Features}

\textbf{Instructor-related representational characteristics:} Tian and  Bourguet~\cite{DBLP:conf/ecce/TianB16} observed that 
the most common 
gestures that instructors used as cues were pointing and extending (extend the arms away from the body with palms up) cues.
Results in evaluating the impact of these types of cues have been shown to increase the learning outcome in most cases~\cite{DBLP:journals/chb/WangLHS20, DBLP:journals/ce/PiZZXYH19, DBLP:journals/jcal/WangPH19, DBLP:journals/bjet/PiHY17}.
However, no significant effects~\cite{DBLP:conf/icls/SharmaDGD16} or even negative effects where found when the instructor was an animation~\cite{DBLP:journals/jche/MoonR21}.

\textbf{Content-related representational characteristics:} Examples of these characteristics are arrows, color-coded cues, highlighted content, etc.
Experiments showed that these cues improve the learning outcome in most cases~\cite{DBLP:journals/chb/WangLHS20, DBLP:journals/bjet/PiHY17, DBLP:conf/icls/SharmaDGD16}, but had no significant effect on cognitive load~\cite{DBLP:conf/icls/SharmaDGD16} or increased it~\cite{DBLP:journals/jche/MoonR21}.
When comparing content-related cues with instructor-related ones, the results were contradictory.
In one case, instructor-related cues had a higher positive impact on the learning outcome than content-related cues when hand-pointing gestures were compared to arrows on slide-based videos~\cite{DBLP:journals/bjet/PiHY17}. 
In another case, contrary results were found 
when comparing hand-pointing gestures (a pencil following the handwriting) with gaze overlay defined by an expert 
(explicit drawing 
of the path that learners should follow with their eyes)~\cite{DBLP:conf/icls/SharmaDGD16}.
    
\textbf{Other visual and audio elements:} General visual and audio features such as the image and audio quality seemed to matter to the learners and caused positive effects on the learning outcome~\cite{DBLP:journals/ijcallt/Ding18, DBLP:conf/sigcse/SchreiberD17}. 
In the experiments of Uchidiuno et al.~\cite{DBLP:journals/aiedu/UchidiunoKHYO18}, English language students often avoided videos that did not have visual support (figures, charts, equations, or on-screen text) but presented only an instructor talking. 
Interestingly, Ding~\cite{DBLP:journals/ijcallt/Ding18} found that music in the background helped to keep students focused.

\subsection{Production Style}

\textbf{Instructor presence:} Experiments have shown that viewers preferred a video in which the instructor appears~\cite{DBLP:journals/mta/Perez-NavarroGC21, DBLP:journals/access/DavilaXSG21}. 
Hew and Lo~\cite{DBLP:journals/ile/HewL20} observed that the three most widely used styles in a sample of YouTube videos in computer science had an instructor:
\begin{inparaenum}[(1)]
    \item slide-based 
    videos,
    \item classroom-style videos, and 
    \item Khan-style videos with talking head.
\end{inparaenum}
In addition, viewers voiced their preference for videos that show the instructor's hands~\cite{DBLP:journals/mta/Perez-NavarroGC21}. 
However, although the presence of an instructor was preferred, it did not bring notable advantages~\cite{DBLP:journals/access/DavilaXSG21, DBLP:journals/ile/HewL20}. 
Davila et al.~\cite{DBLP:journals/access/DavilaXSG21} and Hew and Lo~\cite{DBLP:journals/ile/HewL20} found that a talking head did not show a significant effect on the learning outcome, the social presence~\cite{DBLP:journals/ile/HewL20} and the cognitive load~\cite{DBLP:journals/access/DavilaXSG21}. 
Experiments have also tested the effects of the instructor's size in the video. 
A small size seemed to cause higher learning gains~\cite{DBLP:journals/jcal/PiHY17} and satisfaction~\cite{DBLP:conf/icetc/CaoNW18} when the topic was not complex.
However, the size of the instructor had no significant effect on cognitive load, mental effort~\cite{DBLP:journals/jcal/PiHY17, DBLP:conf/icetc/CaoNW18} and social presence~\cite{DBLP:journals/jcal/PiHY17, DBLP:conf/icetc/CaoNW18}.

\textbf{Dialogue-style vs. Monologue-style:} The experiments did not find a significant effect of dialogue-style over the monologue-style on learning gain~\cite{DBLP:conf/icls/DingASBC18, DBLP:conf/collabtech/NugrahaWZHI20, DBLP:conf/chi/LeeM20}. 
In addition, no effect on the social presence was found~\cite{DBLP:conf/chi/LeeM20}. 
Ohashi et al.~\cite{ohashi2019comparison} conducted an experiment using videos in computer programming topics, one using slides only, another one with slides accompanied by two puppets in a dialogue. 
They found that 
the 
number of viewers of the video including the dialogue decreased over time (throughout a compulsory course). 
The authors argued 
that the puppets' dialogue was effective at the beginning, but not when the learners became increasingly familiar with the course content.
Nevertheless, the dialogue style seemed to offer benefits with regard to self-efficacy~\cite{DBLP:conf/chi/LeeM20} and positive sentiment~\cite{DBLP:conf/icls/DingASBC18}.
    
\textbf{Animations:}
Animations in videos caused higher learning outcomes when compared to 
\begin{inparaenum}[(a)]
    \item text-only videos (i.e., slide-based videos only with text, without illustrations, figures, charts, and similar),~\cite{DBLP:conf/lak/SrivastavaNLVE020}
    \item textbooks~\cite{DBLP:journals/ijicte/CooksonKH20}, and
    \item traditional live classroom lectures~\cite{DBLP:journals/ijim/Al-KhateebA20}.
\end{inparaenum}
Animations also affected other aspects such as participation~\cite{DBLP:journals/ijicte/CooksonKH20}, cognitive load~\cite{DBLP:conf/chi/LeeM20}, mental effort~\cite{DBLP:conf/lak/SrivastavaNLVE020}, and perceived difficulty~\cite{DBLP:conf/chi/SrivastavaVLEB19}.
    
\textbf{Video tutorials and video lectures:} In several studies, learners believed that video tutorials 
were more interesting, attractive, kept them engaged, and helped them understand better~\cite{DBLP:journals/ijcallt/Ding18, DBLP:conf/lats/OuGJH16}.
It was also found that learners tended to skip theoretical discussions, which were usually found in video lectures~\cite{DBLP:journals/ijmlo/LeiYLTYL19}.
However, other studies showed that learners not only watched lectures significantly more often than tutorials~\cite{DBLP:journals/i-jep/HildebrandA18} but also watched them significantly more often to completion~\cite{DBLP:conf/lats/ThorntonRW17}. 
In any case, Hildebrand and Ahn~\cite{DBLP:journals/i-jep/HildebrandA18} reported that both styles had a similar positive correlation with the learning outcome.

\textbf{Other:} Other studies found that media diversity correlated positively with learning outcome, satisfaction, engagement, and interest~\cite{DBLP:journals/itse/CostleyL17}. 
Shoufan and Mohamed~\cite{DBLP:conf/sigite/ShoufanM17} reported that the most predominant styles in CS videos were paper-based explanations, whiteboard-based explanations, slide-based videos, and Khan-style videos.

\subsection{Pace of the Video}

\textbf{Automatic pauses vs. non-automatic pauses:} No significant positive effects of automatic pauses in videos were found compared to non-automatic pauses (where learners can pause the video, also called self-pacing)~\cite{DBLP:journals/tkl/SchroederCC20}. 
Indeed, systems with automatic pauses in videos were ineffective~\cite{DBLP:journals/bjet/Garrett21} or even increased cognitive load~\cite{DBLP:journals/bjet/Garrett21, DBLP:conf/amcis/Garrett18}.
Merkt et al.~\cite{DBLP:journals/chb/MerktBFS18} did not find advantages of automatic pauses even in long lasting videos.
On the contrary, viewers in the 
non-automatic pauses condition were more successful and reported a lower mental effort~\cite{DBLP:conf/amcis/Garrett18, DBLP:journals/tkl/SchroederCC20}. 

Video acceleration: The acceleration rate at which the video is watched was found to be important for learners ~\cite{DBLP:journals/ijcallt/Ding18}. 
Furthermore, there was evidence that students tend to accelerate the videos at least once~\cite{DBLP:journals/elearn/YanB18}.
However, the impact reported on the learning outcome has been diverse. 
On the one hand, Lang et al.~\cite{DBLP:conf/lak/LangCMP20} showed that when the learner accelerated the video, the learning outcome improved for learners in Stanford Online~\cite{veena_stanford_2020} (a platform of open online courses) in diverse disciplines such as CS and mathematics. 
On the other hand, Garrett~\cite{DBLP:journals/bjet/Garrett21} reported that learners who watched 
accelerated CS videos did not outperform those who watched  
non-accelerated videos.
However, in this experiment the learner could not decide when to play, pause or accelerate the video but this was given automatically by the system.

\subsection{Beginners, Specialists, High and Low Performers}

Learners show specific behaviors depending on their performance level with respect to the learning topic.
High achievers engaged in more diverse activity patterns (e.g., play, pause seek)~\cite{DBLP:conf/educon/RodriguezNFPM18}, interacted more intensely with quizzes embedded in the video~\cite{DBLP:conf/lats/Kovacs16}, and 
watched videos more constantly every week~\cite{DBLP:conf/educon/RodriguezNFPM18}.
Galster et al.~\cite{DBLP:conf/icse/GalsterMG18} reported that
students who were more active through in-video annotations, also watched more videos.
One study suggested that if low-performance learners imitate this behavior, advantages could be attained~\cite{DBLP:conf/sigcse/AngraveZHM20}. 
However, Galster et al.~\cite{DBLP:conf/icse/GalsterMG18} did not find differences in learning outcomes between more engaged and less engaged learners.

Regarding the differences between beginners and specialists, the findings showed that specialists engaged more often in different video activities than beginners~\cite{DBLP:journals/ijhci/Li19} and achieved a higher learning outcome~\cite{DBLP:journals/bjet/StohrSMNM19}. 
Nevertheless, the same level of engagement was found in both groups and they also consumed the same amount of videos~\cite{DBLP:journals/bjet/StohrSMNM19, DBLP:journals/ijhci/Li19}.

\subsection{Video Completion and Length}

\textbf{Video completion:} It was reported that learners usually watched 30\% to 50\% of the total video length (on a YouTube channel~\cite{DBLP:journals/ce/SaurabhG19} and on the MOOC platform edX~\cite{DBLP:journals/bjet/StohrSMNM19, edx}) and that skipping through videos was a very common behavior~\cite{DBLP:journals/i-jep/HildebrandA18}.
However, studies in other learning contexts obtained different results. 
If the videos were part of a university course, most of the learners watched the full video~\cite{DBLP:journals/elearn/YanB18, DBLP:journals/bjet/GiannakosJK16}. 
Regarding the gain in the learning outcome, studies showed that the extent to which a video was watched impacted the student's grades~\cite{DBLP:journals/i-jep/HildebrandA18, DBLP:conf/sigcse/AngraveZHM20} and that viewers who watched the video to completion were more likely to watch other videos~\cite{DBLP:conf/amia/WynnWLSBWC19}. 
Regarding dropping behavior, the findings show that this occurred within the first five seconds. 
On the other hand, when the learners watched the video for a longer time, dropping behavior seemed to be reduced~\cite{DBLP:conf/amia/WynnWLSBWC19}. 
Wynn et al. found that when viewers surpassed the first five minutes, they were likely to watch the video to completion~\cite{DBLP:conf/amia/WynnWLSBWC19}.
Strategies such as including in-video quizzes seemed to reduce video dropout~\cite{DBLP:conf/lats/Kovacs16}.

\textbf{Video length:} Several studies have concluded that short learning videos were preferred by learners~\cite{DBLP:journals/ce/SaurabhG19, DBLP:journals/ijcallt/Ding18, rahim2019video, DBLP:conf/emoocs/EngenessN19}. 
Videos within five to seven minutes were preferred in two studies~\cite{DBLP:conf/emoocs/EngenessN19, DBLP:journals/ijcallt/Ding18}. 
One of the studies experimented with videos with talking head, PowerPoint slides, drawings on a board, and animations for learning English as a foreign language by undergraduate students~\cite{ DBLP:journals/ijcallt/Ding18}.
The other study experimented with videos of a MOOC on digital competence that presented pedagogical aspects and tutorials for teachers~\cite{DBLP:conf/emoocs/EngenessN19}.  
Further findings in both studies showed that although longer videos up to 10 minutes were not preferred, they were still well received~\cite{DBLP:conf/emoocs/EngenessN19, DBLP:journals/ijcallt/Ding18}. 
Saurabh and Gautam~\cite{DBLP:journals/ce/SaurabhG19} found that the most popular learning videos lasted 10 to 15 minutes.
Nevertheless, contrary findings presented by Shoufan~\cite{DBLP:journals/ce/Shoufan19} showed that videos perceived as too short were not well received. 
In this study, the videos were selected from YouTube with different liking ratios (number of likes of a video divided by the sum of likes and dislikes), different instructors, different speaking accents, and covering different topics in digital logic design and embedded systems targeted to students of an engineering program. 
For lecture styles, Giannakos et al.~\cite{DBLP:journals/bjet/GiannakosJK16} found that the preferred length of videos in software engineering was close to the traditional 45-minute classroom lectures for university students of a computer science degree.
In any case, the context is an important factor to consider to select an appropriate video length~\cite{DBLP:journals/ce/Shoufan19}.

\subsection{Video popularity}

Saurabh and Gautam~\cite{DBLP:journals/ce/SaurabhG19} found that viewers tend to like videos that helped them to learn and dislike the ones that did not. 
Shoufan and  Mohamed~\cite{DBLP:conf/sigite/ShoufanM17} discovered that the number of likes for learning videos increase
\begin{inparaenum}[(a)]
    \item with higher video resolution, 
    \item with higher talking rates (words per second), 
    \item when English was the native language of the instructor, 
    \item when the production style included paper-based or a whiteboard-based explanations, and
    \item when more than one production style was used.
\end{inparaenum}
Later, Shoufan~\cite{DBLP:journals/ce/Shoufan19}, found in a survey that learners liked a video when they perceived that 
\begin{inparaenum}[(a)]
    \item the explanation seemed appropriate 
    \item the presentation used animations, diverse colors, understandable handwriting, and appropriate usage of supporting technology and software,
    \item the video seemed interesting and easy to focus on,
    \item practical examples were included
    \item the instructor was concise, used a simple vocabulary, and
    \item had a confident and clear voice.
\end{inparaenum}
On the other hand, in the same study, the authors~\cite{DBLP:journals/ce/Shoufan19} found that learners perceived that they disliked a video when the explanation was inappropriate, there were no practical examples, the problem was solved in non-common ways, unneeded details were included, the voice of the instructor seemed dull or annoying, or the content was boring and difficult to focus on.
Additionally, in a different study, Shoufan~\cite{DBLP:journals/chb/Shoufan19} analyzed a survey and found that learners perceived that they disliked a video when they felt like they did not understand the video content, when the technical quality of audio or video was insufficient, when the instructor seemed incompetent, not friendly, or not enthusiastic, or the length of the video was not appropriate.
Curiously, it was found that lengthy videos appeared to generate slightly less dislikes~\cite{DBLP:conf/sigite/ShoufanM17}. 

Lee and Osop~\cite{DBLP:journals/oir/LeeOGK17} detected that popular videos have a high amount of comments with strong negative sentiment. 
Moreover, Lei et al.~\cite{DBLP:journals/ijmlo/LeiYLTYL19} found that popular videos had a high number of views but lower retention rates. 
The authors found that popular videos usually provided a general background, while less popular videos dived deeper into the topic, which might be the reason why they had higher retention rates.
Indeed, it was warned that higher amount of views and comments might not necessarily entail a good video quality~\cite{DBLP:journals/oir/LeeOGK17}.

\section{Conclusions}
\label{sec:conclusion}

\begin{figure}[h]
\centering
\includegraphics[height=7cm]{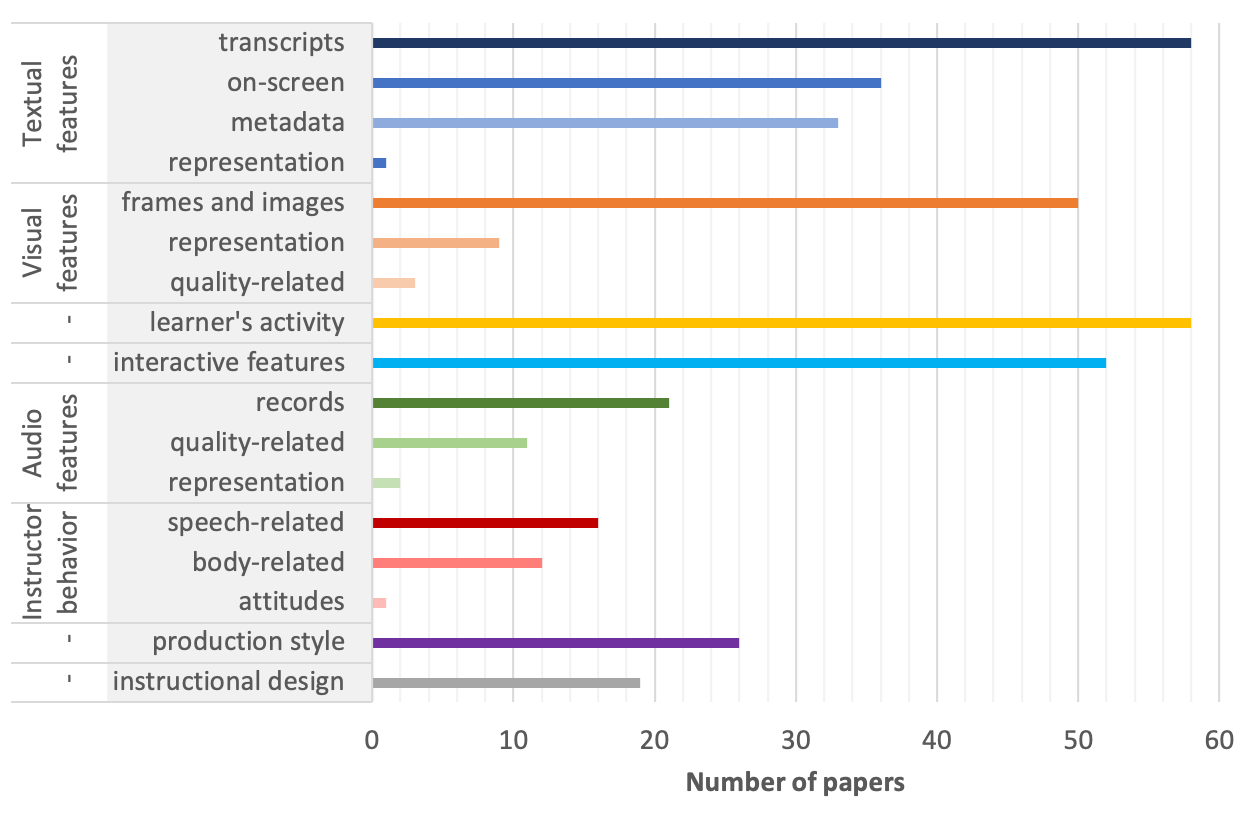}
\caption{the diagram shows all video characteristics identified in this review}
\label{video_features}
\end{figure}

In this paper, we have reviewed 257 articles in the field of video-based learning (2016-2021) following the PRISMA guidelines~\cite{page_prisma_2021}.
Throughout the review, we 
\begin{inparaenum}[(a)]
    \item identified major research directions
    \item surveyed researched video characteristics 
    and organized them into a taxonomy,
    \item reported on target tasks and technologies,
    \item revealed derived trends in instructional design guidelines to produce videos, and
    \item contrasted research findings and summarized  
    the impact of the features on the learning effectiveness. 
\end{inparaenum}

\textbf{Research directions:} We have identified four 
research directions in the field of VBL:
\begin{inparaenum}[(1)]
    \item proposals of tools, 
    \item reports on controlled experiments, 
    \item data analysis studies, and 
    \item design guidelines proposals. 
\end{inparaenum}

The majority of papers proposed tools for the support of VBL, while a moderate proportion of papers presented studies on experiments in controlled settings. A small part of articles were dedicated to data analysis studies and design guidelines.

\textbf{Video characteristics:} Table~\ref{tab:taxonomy} presents the taxonomy we have built based on the investigated characteristics, topics, modalities, etc. identified in the research articles and their corresponding references.
Figure~\ref{video_features} shows a ranking of these video characteristics according to the number of related publications.
The most explored characteristics were textual features followed by visual features, learner activity,
interactive features.
The most used textual features were transcripts, which were mainly employed to develop recommendation systems~(see Section \ref{subsec:recommender systems}) or navigation tools~(See Section \ref{subsec: navigation systems}), as a source to extract information~(see Section \ref{subsec: info extraction}), and for summarization tools~(see Section~\ref{subsec: summarization tools} and Figure~\ref{features_tools}).
The most explored visual features were video frames, 
essentially used to extract information from them.
Characteristics related to the learner activity took advantage of log data and examined, for instance, play and pause activities.
These were heavily explored in data analysis articles~(see Section \ref{subsec: data analysis studies} and Figure \ref{features_research_type}) and in some tool proposals for developing predictors~(see Section \ref{subsec: classifiers}) and analytic tools~(see Section \ref{subsec: analytic tools} and Figure \ref{features_tools}).
Interactive features (e.g., quizzes embedded in the video, in-video annotations) were popularly scrutinized in experiments to evaluate their effectivity~\cite{DBLP:conf/educon/KleftodimosE18, DBLP:journals/ce/AltenPJK20, DBLP:conf/hci/LeisnerZRC20, DBLP:conf/hicss/KellerLFL19,DBLP:conf/lak/ZeeDSGGSPA18, DBLP:journals/ijcses/SozeriK21, DBLP:conf/ectel/TaskinHHDM19,DBLP:conf/cscwd/LuLW21, DBLP:journals/ijicte/KuhailA20, chen2021exploring, DBLP:conf/ecis/WinklerHFSSL20, DBLP:journals/sle/WachtlerHZE16, DBLP:journals/chb/CojeanJ17, DBLP:conf/iiaiaai/KuYC19, DBLP:journals/eait/PalaigeorgiouP19, shelton2016exploring, DBLP:journals/tkl/MoosB16, DBLP:conf/sac/PimentelYMZ19, DBLP:journals/bjet/Chen20, DBLP:conf/ilrn/Munoz-CarpioCB20,DBLP:journals/ce/Araiza-AlbaKMSS21, DBLP:journals/access/DaherS21, DBLP:journals/bjet/HuangHC20, DBLP:conf/hvei/GuervosRPMDG19} but also were integrated in tools for creating videos with interactive elements~(see Section \ref{subsec: interactive tools} Figure~\ref{features_research_type}).

Moderately studied characteristics were related to audio features, production style, and instructor behavior.
Audio features specifically have been used to develop tools~\cite{DBLP:conf/www/MahapatraMR18, DBLP:conf/kes/OhnishiYSNKH19, DBLP:conf/ncc/HusainM19, DBLP:conf/webmedia/SoaresB18, DBLP:journals/prl/KanadjeMAGZL16, DBLP:conf/iui/GandhiBSKLD16, DBLP:conf/iui/KumarSYD17, DBLP:conf/mm/ZhaoLLXW17, DBLP:conf/www/MahapatraMRYAR18, DBLP:conf/intellisys/BianH19, DBLP:journals/jiis/Balasubramanian16, DBLP:conf/fie/WalkYNRG20, DBLP:journals/mta/Furini18, DBLP:journals/cai/GomesDSBS19, DBLP:journals/access/LinLLCWLZL19, DBLP:conf/edm/SharmaBGPD16a, DBLP:conf/mm/ChatbriMLZKKO17, DBLP:conf/ccnc/Furini21a, DBLP:journals/csl/Sanchez-Cortina16, DBLP:conf/ism/GopalakrishnanR17, DBLP:conf/sigir/MukherjeeTC019, DBLP:conf/webmedia/SoaresB18, DBLP:journals/prl/KanadjeMAGZL16, DBLP:conf/cikm/GhauriHE20, DBLP:journals/mta/Furini18, DBLP:conf/ictai/ImranKSK19, DBLP:conf/icrai/ChenWJ20}.
Different types of production style have been tested mainly in controlled experiments~\cite{DBLP:conf/aied/StrancM19, DBLP:conf/icls/DingASBC18, DBLP:journals/ijicte/CooksonKH20, DBLP:journals/ijim/Al-KhateebA20, DBLP:journals/mta/Perez-NavarroGC21, DBLP:conf/collabtech/NugrahaWZHI20, DBLP:conf/chi/LeeM20, DBLP:conf/lak/SrivastavaNLVE020, DBLP:journals/access/DavilaXSG21, DBLP:conf/chi/SrivastavaVLEB19, DBLP:journals/chb/StullFM18, DBLP:journals/jcal/PiHY17, DBLP:journals/ile/HewL20, ohashi2019comparison, DBLP:conf/icetc/CaoNW18} (see Figure~\ref{features_research_type}).
Regarding instructor behavior, speech-related characteristics~\cite{DBLP:journals/chb/HorovitzM21, DBLP:journals/chb/Shoufan19, DBLP:journals/ijcallt/Ding18, DBLP:conf/lats/SluisGZ16, DBLP:conf/lak/AtapattuF17, DBLP:conf/sigite/ShoufanM17, DBLP:conf/cchi/PatelGZ18, DBLP:conf/www/MahapatraMR18, DBLP:conf/webmedia/SoaresB19, DBLP:journals/jiis/Balasubramanian16, DBLP:conf/lats/JoYK19, DBLP:journals/nrhm/JungSKRH19, DBLP:conf/icalt/CheSYM16, DBLP:journals/nrhm/JungSKRH19, DBLP:conf/icuimc/JiKJH18, DBLP:conf/edm/BulathwelaPLYS20} have been used more often than body-related characteristics~\cite{DBLP:journals/ce/BeegeNSNSWMR20, DBLP:journals/chb/BeegeNSR19, DBLP:journals/bjet/WangLCWS19, DBLP:journals/chb/HorovitzM21, DBLP:journals/chb/LiKBJ16, DBLP:journals/ijcallt/Ding18, DBLP:conf/ecce/TianB16, DBLP:conf/icrai/ChenWJ20, DBLP:conf/icpr/XuDSG20, DBLP:conf/bigmm/GodavarthiSMR20, DBLP:journals/chb/HoogerheideWNG18, DBLP:conf/sigite/ShoufanM17} (see Figure~\ref{video_features}).
Less investigated characteristics were instructional design characteristics, which were mainly treated in articles on guidelines for the design and creation of videos~\cite{DBLP:conf/sigcse/MinnesAGF19, DBLP:journals/bjet/WijnkerBGD19, DBLP:conf/icis/WeinertGBB20, DBLP:conf/iceit/GeL20, DBLP:conf/tale/LeiYKLA16, DBLP:conf/icteri/Voronkin19, lange2020improving, DBLP:conf/sigcse/Stephenson19, DBLP:conf/lats/DodsonRFYHF18, hodges2018ensuring, DBLP:journals/ce/HungKC18}.

\textbf{Tools, tasks and technologies:} The main tools, pipelines, frameworks, and methods that have been proposed for supporting VBL were related to information extraction, integration of interactive features, recommendation systems, classification and prediction, video navigation, and summarization~(see Section \ref{subsec: tools} Figure \ref{tool_types}). 
Information extraction focused mainly on retrieving text~\cite{DBLP:journals/access/DavilaXSG21, DBLP:conf/icpr/KotaSDSG20, DBLP:journals/ijdar/KotaDSSG19, DBLP:conf/icdar/XuDSG19, DBLP:journals/mta/LeeYCC17, DBLP:conf/fie/GarberPMALS17, zhong2018fast, DBLP:journals/ijkesdp/KhattabiTB19, DBLP:journals/access/DavilaXSG21, DBLP:journals/ijkesdp/KhattabiTB19, DBLP:journals/ijdar/KotaDSSG19, DBLP:journals/mta/LeeYCC17, DBLP:conf/icfhr/DavilaZ18, DBLP:conf/icdar/XuDSG19, DBLP:conf/fie/GarberPMALS17, zhong2018fast, DBLP:journals/ijkesdp/KhattabiTB19}, 
metadata~\cite{DBLP:journals/nrhm/JungSKRH19, DBLP:conf/icse/Escobar-AvilaPH17, DBLP:journals/cai/GomesDSBS19, DBLP:journals/mta/Furini18, ghosh2021augmenting, DBLP:conf/icalt/NangiKRB19},
and key information from videos such as key video segments and key topics ~\cite{DBLP:conf/cikm/GhauriHE20, DBLP:conf/icalt/CheSYM16, DBLP:journals/mta/KravvarisK19, DBLP:journals/access/LinLLCWLZL19, DBLP:journals/corr/abs-1807-03179, DBLP:journals/tse/PonzanelliBMOPH19}. 
Some approaches used deep learning for information extraction~\cite{DBLP:journals/access/DavilaXSG21, DBLP:journals/ijkesdp/KhattabiTB19, DBLP:conf/icalt/CheSYM16, DBLP:journals/access/LinLLCWLZL19}, but also other (shallow) machine learning methods based on hand-crafted features were still present~\cite{DBLP:conf/icalt/CheSYM16, DBLP:journals/corr/abs-1807-03179, DBLP:journals/tse/PonzanelliBMOPH19}.
Knowledge bases~\cite{DBLP:journals/cai/GomesDSBS19, ghosh2021augmenting, DBLP:conf/icalt/NangiKRB19} and natural language~\cite{DBLP:journals/nrhm/JungSKRH19, DBLP:conf/icse/Escobar-AvilaPH17, DBLP:journals/mta/Furini18} processing methods were also utilized. 

The development of interactive tools mainly targeted the integration of in-video annotations (feature that allow the learner to take notes while watching the video)~\cite{DBLP:conf/jcdl/FongDZRF18, DBLP:conf/chi/LiuYWW19, DBLP:conf/educon/MalchowBM18, lai2018study, DBLP:conf/sigcse/SinghAML16, DBLP:conf/chi/NguyenL16, DBLP:conf/ihci/DebPB17}, quizzes \cite{DBLP:journals/sle/HeraultLMFE18, DBLP:journals/sle/Kohen-VacsMRJ16, DBLP:conf/fie/WalkYNRG20, DBLP:conf/lak/WachtlerKTE16, ma2019automatic, DBLP:journals/ijai/WachtlerE19}, and, interestingly, also feedback to the learner for preventing mind wandering~\cite{DBLP:journals/ile/LinC19, DBLP:conf/csee/JoL17, DBLP:conf/ht/RobalZLH18, DBLP:conf/lak/SharmaAJD16, DBLP:conf/iciit/Llanda19}. 
A few proposals introduced tools to create videos with 360-degree characteristics~\cite{DBLP:journals/sle/HeraultLMFE18, DBLP:conf/vrst/KoumaditisC19, DBLP:conf/teem/BernsMRD18}.

Recommendation systems have mainly used the similarity between transcripts and metadata to suggest videos~\cite{DBLP:conf/chi/ZhaoBCS18, DBLP:journals/jois/WaykarB17, DBLP:conf/ideal/BleoancaHPJM20, DBLP:conf/mmm/BasuYSZ16, DBLP:journals/nrhm/JungSKRH19, DBLP:conf/icuimc/JiKJH18, DBLP:conf/paams/JordanVTB20, DBLP:conf/fie/WalkYNRG20, DBLP:conf/icalt/TavakoliHEK20}. 
A popular approach was to extract topics, concepts, and keywords to compute the similarity to other videos.
The use of ontologies and external knowledge bases was used particularly to extract concepts~\cite{DBLP:conf/ectel/SchultenMLH20, DBLP:conf/icalt/BorgesR19, DBLP:conf/dms/CoccoliV16, DBLP:journals/access/BorgesS19}.
Some interesting approaches considered other characteristics such as video readability~\cite{DBLP:journals/nrhm/JungSKRH19}, the skills demanded by the labor market~\cite{DBLP:conf/icalt/TavakoliHEK20}, video popularity~\cite{DBLP:journals/snam/KravvarisK17, DBLP:conf/icccsec/ZhangZWL18}, videos from the same instructor~\cite{DBLP:conf/iftc/GuanLMA19, DBLP:conf/ictai/ImranKSK19}, learning style~\cite{DBLP:conf/softcom/CiurezMGHPJ19}, and pedagogical characteristics~\cite{DBLP:journals/or/Acuna-SotoLP20}.

For classifiers and predictors the main task was to predict the learning outcome using 
learner's activity data ~\cite{DBLP:journals/eait/LemayD20, DBLP:conf/aied/MbouzaoDS20, DBLP:journals/eait/MubarakCA21, DBLP:conf/iiaiaai/FurukawaIY20, DBLP:conf/cits/KorosiEFT18, DBLP:conf/aied/LalleC20, DBLP:journals/tsp/BrintonBCP16}. 
Other, but less frequent tasks were discipline classification~\cite{DBLP:conf/hais/StoicaHPJM19, DBLP:conf/esws/DessiFMR17, DBLP:conf/cbmi/BasuYZ16, DBLP:conf/bdca/OthmanAJ17, DBLP:conf/mm/ChatbriMLZKKO17}, instructor behavior prediction ~\cite{DBLP:conf/icrai/ChenWJ20, DBLP:conf/bigmm/GodavarthiSMR20, DBLP:conf/edm/SharmaBGPD16a, DBLP:conf/icpr/XuDSG20}, and dropout prediction ~\cite{DBLP:journals/corr/abs-2002-01955, DBLP:conf/goodit/FuriniGM21}.

Navigation tools popularly proposed table content navigation styles~\cite{DBLP:journals/corr/abs-2012-07589, DBLP:journals/mta/KravvarisK19, 
DBLP:conf/www/MahapatraMR18, DBLP:conf/sigir/MukherjeeTC019, DBLP:conf/ncc/HusainM19, DBLP:conf/iui/GandhiBSKLD16, DBLP:conf/iui/KumarSYD17, DBLP:conf/mm/ZhaoLLXW17, DBLP:conf/iui/YadavGBSSD16, DBLP:conf/intellisys/BianH19} but some interesting approaches have emerged such as navigation according to keywords~\cite{DBLP:conf/kes/OhnishiYSNKH19, DBLP:conf/ism/KokaCRSS20, DBLP:journals/mta/KravvarisK19, DBLP:journals/prl/KanadjeMAGZL16, DBLP:conf/iui/KumarSYD17, DBLP:conf/mm/ZhaoLLXW17} or visual elements~\cite{DBLP:journals/nca/ZhaoLQWL18, DBLP:journals/corr/abs-1712-00575, DBLP:conf/iui/KumarSYD17, DBLP:conf/mm/ZhaoLLXW17, DBLP:conf/iui/YadavGBSSD16}. 
These tools have relied heavily in technologies to segment the video and to extract subtopics, concepts, and keywords.

Analytic tools relied usually 
on characteristics of the learner's activity 
to understand video usage~\cite{DBLP:conf/fie/WalkYNRG20, DBLP:conf/lak/WachtlerKTE16, DBLP:conf/fie/LongTS18} and contrasted it with the learners performance~\cite{DBLP:journals/jcp/MinCX19, DBLP:conf/ieeevast/HeZD18, DBLP:conf/icalt/HeZDY18}.
Some interesting tools analyzed the comments of videos to understand the learners' sentiment~\cite{DBLP:conf/chi/SungHSCLW16} and their engagement~\cite{DBLP:conf/aied/StepanekD17}.

Summarization methods used commonly text transcripts~\cite{DBLP:conf/t4e/VimalakshaVK19, DBLP:journals/jiis/Balasubramanian16, DBLP:journals/jasis/KimK16}, however, there were also approaches using (visual) frame-based video content~\cite{DBLP:conf/ism/RahmanSS20}. 
Relevant methods were related to the extraction of keywords, relevant subtitles, entities, and metadata.

\textbf{Design guidelines:} Some of these studies proposed general guidelines~\cite{DBLP:journals/bjet/WijnkerBGD19, DBLP:conf/icteri/Voronkin19, lange2020improving, DBLP:conf/lats/DodsonRFYHF18}, other suggested guidelines for specific disciplines~\cite{DBLP:conf/sigcse/MinnesAGF19, hodges2018ensuring, DBLP:conf/sigcse/Stephenson19, DBLP:conf/tale/LeiYKLA16, DBLP:conf/iceit/GeL20}, and two cases suggested guidelines for creating interactive videos~\cite{DBLP:conf/icis/WeinertGBB20, DBLP:journals/ce/HungKC18} . 
They have leveraged on theories and frameworks, for example, film theory~\cite{bordwell1993film}, limited capacity theory in multimedia learning~\cite{mayer2014multimedia}, interaction theory~\cite{moore1989three}, active viewing framework~\cite{DBLP:conf/lats/DodsonRFYHF18}, etc. 
These guidelines aimed at the improvement of a wide variety of aspects such as pace, intelligibility, quality, ergonomics, interaction, visuals, etc.

\textbf{Experiments, data analysis, and their results on the effectiveness of video characteristics:} One of the major characteristics explored in controlled experiments were interactive features. 
Other characteristics that were experimented are related to production style, instructor behavior, and cues. 
Learner's activity characteristics have been specially inspected in data analysis studies.

Interactive features focused mainly on quizzes, virtual reality and 360-degree characteristics, and in-video annotations.
Quizzes showed to be effective~\cite{DBLP:conf/educon/KleftodimosE18, DBLP:conf/hci/LeisnerZRC20, DBLP:journals/sle/WachtlerHZE16, DBLP:journals/eait/PalaigeorgiouP19, DBLP:journals/ijcallt/Ding18, DBLP:conf/sac/PimentelYMZ19, DBLP:conf/lats/Kovacs16}, 
but not in all cases -- sometimes they were only effective when they were used together with other interactive features and placed at the right moment~\cite{DBLP:journals/sle/WachtlerHZE16, DBLP:journals/eait/PalaigeorgiouP19, DBLP:journals/ijcallt/Ding18}.
Virtual reality and 360-degree characteristics seemed to be effective; however, the impact was not significant~\cite{DBLP:journals/bjet/Chen20, DBLP:journals/ce/Araiza-AlbaKMSS21, DBLP:journals/access/DaherS21, DBLP:journals/bjet/HuangHC20}.
In-video annotations seemed to improve the learning outcome~\cite{DBLP:conf/hci/LeisnerZRC20,DBLP:conf/ectel/TaskinHHDM19,DBLP:journals/ijicte/KuhailA20} and also navigational elements, specifically, table content navigation and timelines with clickable points, impacted positively the learning outcome~\cite{DBLP:journals/ijicte/KuhailA20, DBLP:conf/sac/PimentelYMZ19, DBLP:journals/chb/CojeanJ17}; however, results were not always statistically significant~\cite{DBLP:conf/sac/PimentelYMZ19} or they were more beneficial when they were together with other features~\cite{DBLP:journals/chb/CojeanJ17}.
Indeed, further claims state that videos with interactive features need to consider other pedagogical techniques to improve the learning outcome~\cite{DBLP:journals/caee/GranjoR18}.

Experiments on production styles investigated mainly the videos with instructor's presence, videos with animations, video tutorials vs. video lectures, and dialogue-style vs. monologue-style. 
The learners preferred the presence of an instructor, although the benefits were not always significant~\cite{DBLP:journals/mta/Perez-NavarroGC21, DBLP:journals/access/DavilaXSG21, DBLP:journals/ile/HewL20, DBLP:journals/mta/Perez-NavarroGC21, DBLP:journals/access/DavilaXSG21, DBLP:journals/ile/HewL20}.
Animations caused higher learning outcomes when compared to text media and traditional class~\cite{DBLP:conf/lak/SrivastavaNLVE020, DBLP:journals/ijicte/CooksonKH20, DBLP:journals/ijim/Al-KhateebA20}.
Tutorials and lecture styles correlated positively with the learning outcome~\cite{DBLP:journals/i-jep/HildebrandA18}, however, preferences for one style over the other were different between studies~\cite{DBLP:journals/ijcallt/Ding18, DBLP:conf/lats/OuGJH16, DBLP:journals/ijmlo/LeiYLTYL19, DBLP:journals/i-jep/HildebrandA18, DBLP:conf/lats/ThorntonRW17}. 
Dialogue-style instead of the typical monologue-style showed no significant effects on learning outcome~\cite{DBLP:conf/icls/DingASBC18, DBLP:conf/collabtech/NugrahaWZHI20, DBLP:conf/chi/LeeM20}. 
In any case, the use of divers styles appeared to be effective~\cite{DBLP:journals/itse/CostleyL17}.

Studies on instructor behavior examined particularly human gestures and speech-related characteristics, but very little other visual elements such as pose and gender.
Speech characteristics seemed to impact certain aspects of the learning effectiveness such as understanding, attention, etc \cite{DBLP:conf/lak/AtapattuF17, DBLP:journals/chb/Shoufan19, DBLP:conf/lats/SluisGZ16}.
Gestures proved to benefit only certain aspects such as satisfaction and presence, but very little the learning outcome~\cite{DBLP:journals/ce/BeegeNSNSWMR20, DBLP:journals/bjet/WangLCWS19, DBLP:journals/chb/HorovitzM21}.
Instructor-related cues caused a positive effect on the learning outcome~\cite{DBLP:journals/chb/WangLHS20, DBLP:journals/ce/PiZZXYH19}. 
However, this did not apply to 
animated instructors (human-like artificial agents), in those it even caused negative effects~\cite{DBLP:conf/icls/SharmaDGD16, DBLP:journals/jche/MoonR21}.
Content-related cues have also positively affected the learning outcome~\cite{DBLP:journals/chb/WangLHS20, DBLP:journals/bjet/PiHY17, DBLP:conf/icls/SharmaDGD16}; however, they can also cause cognitive overload~\cite{DBLP:journals/jche/MoonR21}.

Regarding the choice of video length, several studies found that short videos were preferred by the learners~\cite{DBLP:journals/ce/SaurabhG19, DBLP:journals/ijcallt/Ding18, rahim2019video, DBLP:conf/emoocs/EngenessN19}.
However, this was not true for every type of learning videos and context, in some cases longer videos were still preferred~\cite{DBLP:journals/bjet/GiannakosJK16, DBLP:journals/ce/Shoufan19}.
Regarding video completion, experiments showed that the learning outcome improved when learners watched the video more than a certain proportion~\cite{DBLP:journals/i-jep/HildebrandA18,  DBLP:conf/sigcse/AngraveZHM20}. 
However, video completion was affected by common behavior such as skipping and watching only parts of the video~\cite{DBLP:journals/i-jep/HildebrandA18, DBLP:journals/elearn/YanB18, DBLP:journals/bjet/GiannakosJK16}.
Studies on the impact of video pace showed no significant positive effects of automatic pauses, and even negative effects on the learning outcome were found~\cite{DBLP:journals/tkl/SchroederCC20, DBLP:conf/amcis/Garrett18, DBLP:journals/bjet/Garrett21, DBLP:journals/chb/MerktBFS18}. 
Interestingly, watching videos at faster speed was beneficial in some cases and counterproductive in others~\cite{DBLP:conf/lak/LangCMP20, DBLP:journals/bjet/Garrett21}.

Studies on experts and high performance learners revealed that these type of viewers engaged more actively with the video~\cite{DBLP:conf/educon/RodriguezNFPM18, DBLP:conf/lats/Kovacs16, DBLP:conf/educon/RodriguezNFPM18, DBLP:journals/ijhci/Li19} and achieved a higher learning outcome~\cite{DBLP:journals/bjet/StohrSMNM19}.
However, in some cases they achieved the same learning outcome and engagement as their counterparts~\cite{DBLP:conf/icse/GalsterMG18, DBLP:journals/bjet/StohrSMNM19, DBLP:journals/ijhci/Li19}.
Also, behavior might change according to the discipline, topics, and content of the video~\cite{DBLP:journals/ce/SaurabhG19, DBLP:journals/aiedu/UchidiunoKHYO18, DBLP:journals/i-jep/HildebrandA18}.
Also, YouTube videos might be preferred over institutional videos~\cite{DBLP:journals/bjet/GiannakosJK16}.

Regarding video popularity, studies identified that learners tended to like videos perceived as helpful for learning~\cite{DBLP:journals/ce/SaurabhG19}, however, a high popularity might not necessarily entail a good video quality~\cite{DBLP:journals/oir/LeeOGK17}.

\textbf{Further trends and directions:} Our analyses have confirmed several findings from previous reviews.
We observed an increase in the number of VBL research articles~\cite{DBLP:journals/bjet/Giannakos13, DBLP:conf/lak/PoquetLMD18},
and a significant emphasis on the use of videos in STEM disciplines~\cite{DBLP:journals/bjet/Giannakos13, DBLP:conf/lak/PoquetLMD18}.
We also identified a tendency to use primarily the learning outcome as an indicator to measure learning effectiveness, traditionally in terms of recall tests and, more recently, through transfer tests~\cite{DBLP:conf/lak/PoquetLMD18, yousef2014state}. 
Moreover, similarly to  Poquet et al.~\cite{DBLP:conf/lak/PoquetLMD18}, we found that the next most popular indicators were cognitive load, satisfaction, mental effort, social presence, motivation, and self-efficacy. 
We could also confirm that eye-tracking~\cite{DBLP:conf/collabtech/NugrahaWZHI20, DBLP:conf/chi/LeeM20, DBLP:journals/chb/StullFM18, DBLP:journals/jcal/WangPH19, DBLP:conf/icchp/KikusawaOKR16, DBLP:journals/ce/PiZZXYH19, DBLP:conf/hci/Zhang21, DBLP:conf/icls/SharmaDGD16, DBLP:journals/jche/MoonR21, DBLP:journals/bjet/PiHY17, DBLP:conf/chi/SrivastavaVLEB19} and log data~\cite{DBLP:conf/sac/PimentelYMZ19, DBLP:conf/lak/LangCMP20} characteristics were used; however, we also found studies that used thermal \cite{DBLP:conf/chi/SrivastavaVLEB19, DBLP:conf/lak/SrivastavaNLVE020}, neurophysiological \cite{DBLP:conf/ecis/WinklerHFSSL20}, and facial expressions-tracking~\cite{DBLP:conf/chi/SrivastavaVLEB19} characteristics.
Regarding production styles, we could recognize slide-based productions as one of the most used production styles in controlled experiments.
In addition, we found that the experiments frequently used videos with a visible instructor. 
However, contrary to Poquet et al.~\cite{DBLP:conf/lak/PoquetLMD18}, 
we did not find that animations were popularly used.
Among the styles that were less used, there were handwriting, screencast, and virtual reality styles.
We conclude that VBL research in controlled experiments have not tested the diverse range of learning video styles, disciplines, and video sources. 
Thus, the field requires more research efforts in the areas not targeted yet.

Another tendency in the reviewed articles was that audio, visual, textual and further features are usually treated in a separated manner.
However, learning through videos requires the fusion and interpretation of synchronous audio and visual sensory input. 
Thus, we believe that VBL could benefit from 
multimodal representations and analysis 
in order to better capture the relationships between the different modalities.

Research on video, text, and image analysis has been dominated by deep learning approaches in recent years.
This is also the case for VBL, 
especially for the development of tools that support the learning process. 
We noticed that deep learning was not only used for basic tasks such as video segmentation, optical character recognition, neural word representations (embeddings), speech-to-text tasks, etc. but also in entire pipelines of recommendation systems, learning outcome predictors, navigational tools, etc.

Contradictory results were found about the impact of video characteristics on the learning effectiveness.
We could observe that the discrepancies originate not only from the small or big differences in the test conditions but also from the production style of the video.
In some cases it was not possible to know in detail these differences, since not all the studies reported accurately the characteristics of the investigated videos.

\textbf{Future work:} 
Experimental research should consider more the diversity in video styles and also the diverse disciplines apart from STEM subjects.
An important gap we see is that research needs to move towards multimodal representations were video characteristics can be unified to approach to the way human beings learn. 
There are very few articles that propose design guidelines. 
We believe that more research in this matter, especially guidelines customized to specific disciplines, can be highly useful for practical scenarios such as content creation.
Although articles that propose tools for supporting learning encompass most of VBL research in our sample, we think that more efforts are needed to bring new technologies to the field for understanding whether they can improve the way people learn through videos.
Finally, finding what characteristics a video should have to be effective for learning remains ongoing work. 
Obviously, more research is needed to identify what makes us learn in a video and in what learning scenarios. 

\textbf{Limitations:} In this review, we searched and retrieved articles exclusively from academic computer science data bases.
Therefore, we did not consider other disciplines that could contribute to VBL advances, e.g., instructional insights from science education.
We have researched in three dedicated databases, where one of them (DBLP) can be considered to track the vast majority of relevant CS venues, including journals and conference proceedings~\cite{dblp}. 
We retrieved articles with the query terms only in the title to make sure that the main focus was on VBL -- however, there might be articles treating VBL-related topics without it being reflected in the title.
Moreover, we restricted our review to articles that used and experimented with video characteristics, thus, advances in other use cases such as teacher reflection were not reported.
Finally, it is important to recognize that an accurate comparison of the effects of video characteristics on the learning effectiveness is difficult due to the heterogeneity in the experimental settings and test conditions.
This is the reason why we could not provide a meta-analysis about these effects.

\begin{acks}

This work has been partly supported by the Ministry of Science and  Education of Lower Saxony, Germany, through the PhD Training Program ``LernMINT: Data-assisted classroom teaching in the MINT subjects''. 

\end{acks}

\bibliographystyle{ACM-Reference-Format}
\bibliography{manuscript}

\end{document}